\documentclass{egpubl}
\usepackage{egsr17}

 \JournalPaper         
 \electronicVersion 

\ifpdf \usepackage[pdftex]{graphicx} \pdfcompresslevel=9

\else \usepackage[dvips]{graphicx} \fi

\PrintedOrElectronic

\usepackage{t1enc,dfadobe}

\usepackage{egweblnk}
\usepackage{cite}
\usepackage{booktabs}

\usepackage[utf8]{inputenc}                 
\usepackage{fontenc}                        
\usepackage{graphicx}   
\usepackage{float}      
\usepackage{subcaption} 
\usepackage{import}     
\usepackage{mathtools} 
\usepackage{amssymb}   
\usepackage{nicefrac}  
\usepackage{relsize}   
\usepackage{ifmtarg}   
\usepackage{xparse}    
\usepackage{pgffor}    
\usepackage{times} 
\usepackage{microtype}

\usepackage{pstricks}

\usepackage{color}
\usepackage[plain]{fancyref}

\definecolor{gray}{rgb}{0.8,0.8,0.8}
\definecolor{cyan}{cmyk}{1,0,0,0}
\definecolor{blue}{rgb}{0,0,0.7}
\definecolor{red}{rgb}{0.5,0.0,0}
\definecolor{darkgreen}{rgb}{0,0.5,0}
\definecolor{orange}{rgb}{1,0.5,0}
\definecolor{magenta}{cmyk}{0,1,0,0}
\definecolor{darkyellow}{cmyk}{0,0,0.75,0}
\definecolor{purple}{cmyk}{0.5,1,0,0}
\definecolor{darkorange}{rgb}{0.7,0.4,0}
\definecolor{lightred}{rgb}{0.9,0.0,0}
\definecolor{lightblue}{rgb}{0.0,0.0,0.9}
\definecolor{lightgreen}{rgb}{0.0,0.7,0.0}

\newcommand{\new}[1]{#1}

\newcommand{\citet}[1]{\cite{#1}}

\newcommand{\Figs}[2]{Figures~\ref{#1} and~\ref{#2}}
\newcommand{\Eqs}[2]{Equations~\eqref{#1} and~\eqref{#2}}

\newcommand*{\fancyrefapplabelprefix}{app}

\fancyrefaddcaptions{english}{%
\providecommand*{\frefappname}{appendix}%
\providecommand*{\Frefappname}{Appendix}%
}

\frefformat{plain}{\fancyrefapplabelprefix}{%
\Frefappname\fancyrefdefaultspacing#1%
}
\Frefformat{plain}{\fancyrefapplabelprefix}{%
\Frefappname\fancyrefdefaultspacing#1%
}

\frefformat{vario}{\fancyrefapplabelprefix}{%
\frefappname\fancyrefdefaultspacing#1#3%
}
\Frefformat{vario}{\fancyrefapplabelprefix}{%
\Frefappname\fancyrefdefaultspacing#1#3%
}

\makeatletter
\newcommand{\isempty}[3]{%
	\@ifmtarg{#1}
		{#2} 
		{#3} 
}
\makeatother

\makeatletter
\newcommand{\isnotempty}[2]{%
	\@ifmtarg{#1}
		{}
		{#2}
}
\makeatother

\newcommand{\unaryminus}{\scalebox{0.5}[1.0]{\( - \)}}

\newcommand{\slfrac}[2]{\left.#1\middle/#2\right.}

\DeclareDocumentCommand{\indicator}{ o m }{
	\ensuremath{
		\mathlarger{\operatorname{\chi}}
		\IfNoValueTF{#1}{}
			{_{#1}}
		\IfNoValueTF{#2}{}
			{\isnotempty{#2}{\left(#2\right)}}
	}
}

\newcommand{\point}[1]{\ensuremath{\mathbf{#1}}}
\newcommand{\normal}[1]{\ensuremath{\widehat{#1}}}

\newcommand{\refSystem}[1]{\mathcal{R}_{\mathrm{#1}}}
\DeclareDocumentCommand{\axis}{ m o }{
	\ensuremath{
		\normal{#1}\IfValueT{#2}{_{\mathrm{#2}}}
	}
}

\DeclareDocumentCommand{\function}{ m }{
	\ensuremath{
		f \IfValueT{#1}{\isnotempty{#1}{\left(#1\right)}}
	}
}

\newcommand{\diff}{\ensuremath{\mathop{}\!\mathrm{d}}}
\DeclareDocumentCommand{\measure}{ m }{
	\ensuremath{
		\mu\IfValueT{#1}{\isnotempty{#1}{\left(#1\right)}}
	}
}

\DeclareDocumentCommand{\firstKind}{ o m }{
	\ensuremath{
		\IfValueTF{#1}{\operatorname{F}\left( #1|}{\operatorname{K}\left(} #2 \right)
	}
}

\DeclareDocumentCommand{\secondKind}{ o m }{
	\ensuremath{
		\operatorname{E}\left( \IfValueT{#1}{#1|} #2 \right)
	}
}

\DeclareDocumentCommand{\thirdKind}{ m o m }{
	\ensuremath{
		\operatorname{\Pi}\left( #1; \IfValueT{#2}{#2|} #3 \right)
	}
}

\DeclareDocumentCommand{\bsdf}{ > {\SplitArgument{2}{,}} m }{
	\bsdfInternal #1
}

\DeclareDocumentCommand{\bsdfInternal}{ m m m }{
	\ensuremath{
		\rho\IfValueT{#1}{\isnotempty{#1}{
				\left(#1,#2\IfValueT{#3}{,#3}\right)
		}}
	}
}

\DeclareDocumentCommand{\areaMeasure}{ m }{
	\ensuremath{
		\mu_{\mathcal{A}}\IfValueT{#1}{\isnotempty{#1}{\left(#1\right)}}
	}
}

\newcommand{\solidAngle}{\ensuremath{\Omega}}
\newcommand{\lightSolidAngle}{\ensuremath{\Omega_D}}

\DeclareDocumentCommand{\saMeasure}{ m }{
	\ensuremath{
		\mu_{\solidAngle}\IfValueT{#1}{\isnotempty{#1}{\left(#1\right)}}
	}
}

\DeclareDocumentCommand{\pdf}{ o m }{
	\ensuremath{
		p\IfValueT{#1}{^{#1}}\IfValueT{#2}{\isnotempty{#2}{\left(#2\right)}}
	}
}

\DeclareDocumentCommand{\CDF}{ o m }{
	\ensuremath{
		P\IfValueT{#1}{^{#1}}\IfValueT{#2}{\isnotempty{#2}{\left(#2\right)}}
	}
}

\DeclareDocumentCommand{\MSE}{ m }{
	\ensuremath{
		\operatorname{MSE}\IfValueT{#1}{\isnotempty{#1}{\left[#1\right]}}
	}
}

\DeclareDocumentCommand{\RMSE}{ m }{
	\ensuremath{
		\operatorname{RMSE}\IfValueT{#1}{\isnotempty{#1}{\left[#1\right]}}
	}
}

\newcommand{\variance}[1]{\ensuremath{\sigma}}

\DeclareDocumentCommand{\Estimate}{ m o }{
	\ensuremath{
		\widehat{#1}\IfValueT{#2}{_{#2}}
	}
}

\DeclareDocumentCommand{\randVar}{ o }{
	\ensuremath{
		\varepsilon\IfValueT{#1}{_{#1}}
	}
}

\newcommand{\atang}{\ensuremath{a_\mathrm{t}}}
\newcommand{\btang}{\ensuremath{b_\mathrm{t}}}
\newcommand{\ctang}{\ensuremath{c_\mathrm{t}}}

\newcommand{\angleRadial}{\ensuremath{\phi_\mathrm{r}}}
\newcommand{\anglePolar}{\ensuremath{\phi_\mathrm{p}}}
\newcommand{\hRadial}{\ensuremath{h_\mathrm{r}}}
\newcommand{\hPolar}{\ensuremath{h_\mathrm{p}}}
\newcommand{\solidAngleRadial}{\ensuremath{\solidAngle_\mathrm{r}}}
\newcommand{\solidAnglePolar}{\ensuremath{\solidAngle_\mathrm{p}}}
\newcommand{\solidAnglePolarPos}{\ensuremath{\solidAngle_\mathrm{p}^{\mathrm{\tiny +}}}}

\newcommand{\px}{\ensuremath{\point{o}}}
\newcommand{\py}{\ensuremath{\point{x}}}

\newcommand{\wo}{\ensuremath{\normal{\omega}_\mathrm{o}}}

\newcommand{\ww}{\ensuremath{\normal{\omega}}}

\newcommand{\contrib}{f}
\newcommand{\brdf}{f_\mathrm{s}}
\newcommand{\Ls}{L_\mathrm{s}}
\newcommand{\Le}{L_\mathrm{e}}
\newcommand{\shortminus}{\scalebox{0.66}[1.0]{$-$}}

\title[Area-preserving parameterizations for spherical ellipses]%
      {Area-Preserving Parameterizations for Spherical Ellipses}

\author[Guill\'{e}n et al.]
{\parbox{\textwidth}{\centering
		Ib\'{o}n Guill\'{e}n$^{1,2}$
		\quad
		Carlos Ure\~{n}a$^{3}$
		\quad
		Alan King$^{4}$
		\quad
		Marcos Fajardo$^{4}$
		\quad
		Iliyan Georgiev$^{4}$
		\quad
		Jorge L\'{o}pez-Moreno$^{2}$
		\quad
		Adrian Jarabo$^{1}$
	}
	\\
{\parbox{\textwidth}{\centering
		$^1$Universidad de Zaragoza, I3A \quad
		$^2$Universidad Rey Juan Carlos \quad
		$^3$Universidad de Granada \quad
		$^4$Solid Angle
	}
}}

\volume{36}   
\issue{4}     
\pStartPage{179}      

\begin{document}


\maketitle

\begin{abstract}

We present new methods for uniformly sampling the solid angle subtended by a disk.
To achieve this, we devise two novel area-preserving mappings from the unit
square $[0,1]^2$ to a spherical ellipse (i.e.\ the projection of the disk onto the
unit sphere). These mappings allow for low-variance stratified sampling of
direct illumination from disk-shaped light sources. We discuss how to efficiently
incorporate our methods into a production renderer and demonstrate the quality of
our maps, showing significantly lower variance than previous work.

\ccsdesc[500]{Computing methodologies~Rendering}
\ccsdesc[500]{Computing methodologies~Ray tracing}
\ccsdesc[300]{Computing methodologies~Visibility}

\printccsdesc   
\end{abstract}

\section{Introduction}
\label{sec:intro}

Illumination from area light sources is among the most important lighting effects
in realistic rendering, due to the ubiquity of such sources in
real-world scenes. Monte Carlo integration is the standard method for
computing the illumination from such luminaires~\cite{Shirley:1996:Direct}. This
method is general and robust, supports arbitrary reflectance models and geometry, and
predictively converges to the actual solution as the number of samples increases.
Accurately sampling the illumination from area light sources is crucial for minimizing
the amount of noise in rendered images.

Estimating the direct illumination at a point requires sampling the radiance 
contribution from directions inside the solid angle subtended by the given luminaire. A 
sensible strategy is to distribute those directions uniformly. This, however, is hard 
to achieve for an arbitrary-shaped luminaire, as it involves first computing and then uniformly sampling its subtended solid angle.
Specialized methods have been proposed for
spherical~\cite{Wang:1992:Direct},
triangular~\cite{Arvo:1995:Triangles, Urena:2000:Triangles},
rectangular~\cite{Urena:2013:Quads}, and
polygonal lights~\cite{Arvo:2001:Manifolds}.
These elaborate solid angle sampling techniques are more computationally expensive than na\"ive
methods that uniformly sample the surface area of the luminaire. However,
in most non-trivial scenes, where the sample contribution evaluation is orders of
magnitude more costly than the sample generation, their lower variance improves overall efficiency.

Few papers have focused on sampling oriented disk-shaped light sources. Disk lights
are important in practice, both for their artistic expressiveness
and their use in a number of real-world scenarios, generally including
man-made light sources such as in architectural lighting, film and photography.
Moreover, disk lights form the base for some approximate global illumination
algorithms~\cite{Hasan:2009:VSL, Simon:2015:Rich}.
So far, the only practical method for uniformly sampling the solid angle of disk lights is
the work by Gamito~\cite{Gamito:2016:Disks}, who proposed a rejection sampling approach that generates candidates using spherical quad sampling~\cite{Urena:2013:Quads}. Unfortunately, achieving good sample stratification with this method requires special care.

In this paper we present a set of 
methods for uniformly sampling the solid angle subtended by an
oriented disk. We exploit the fact that a disk, as seen from a point, is
bounded by an elliptical cone~\cite{Eberly:1999:Ellipses} and thus its solid angle defines
a spherical ellipse whose properties have been analyzed in
depth~\cite{Booth:1844:SphericalEllipse}. This allows us to define two different
exact area-preserving mappings that can be used to transform stratified unit-square
sample patterns to stratified directions on the subtended spherical ellipse.
We describe how to efficiently implement these mappings in practice and demonstrate the lower variance they achieve compared to previous work.


\newcommand{\fithlinewidth}{.18\linewidth}
\begin{figure*}[!ht]
		\centering
		\def\svgwidth{\fithlinewidth}
		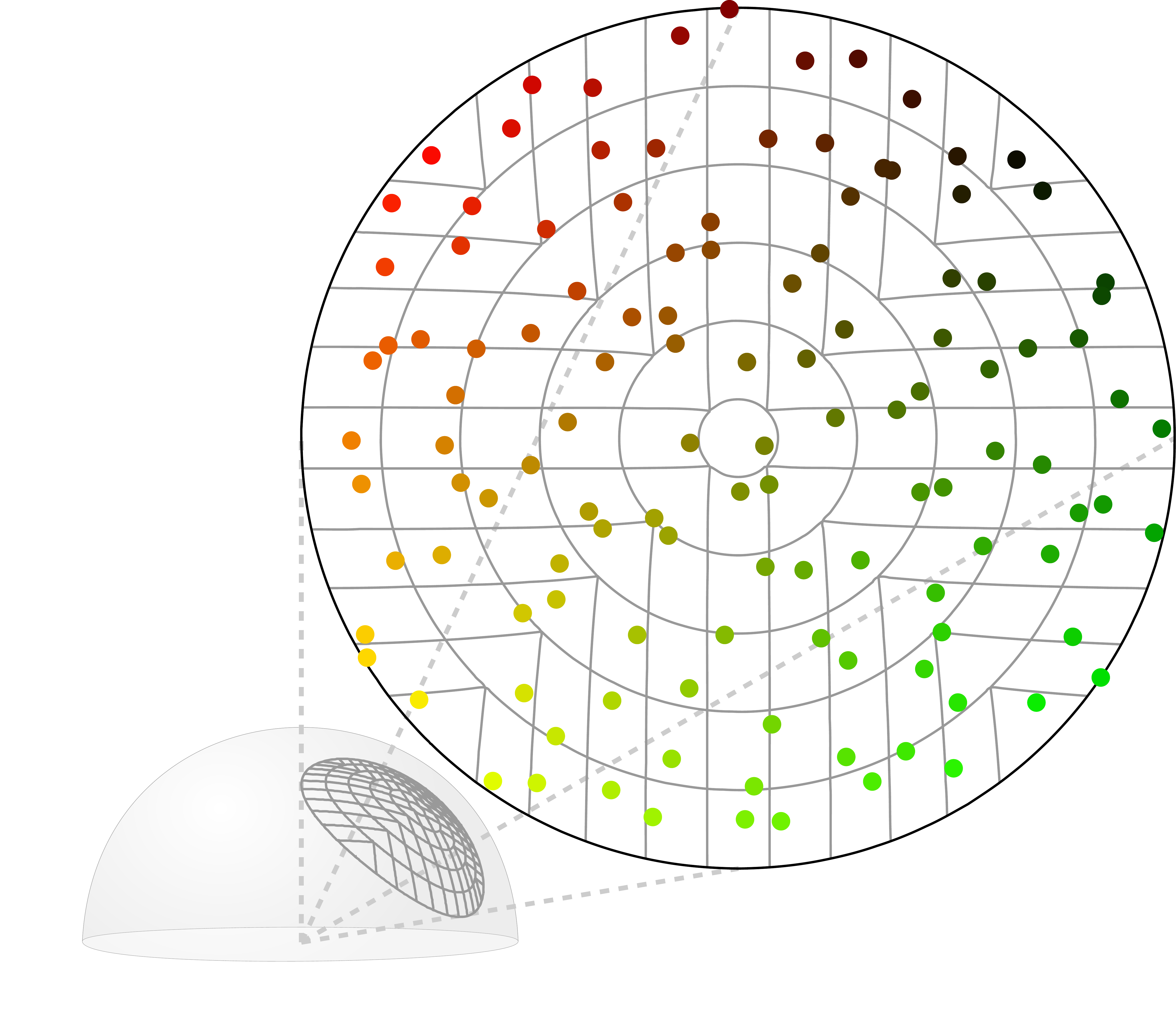
		\hspace{0.26cm}
		\def\svgwidth{\fithlinewidth}
		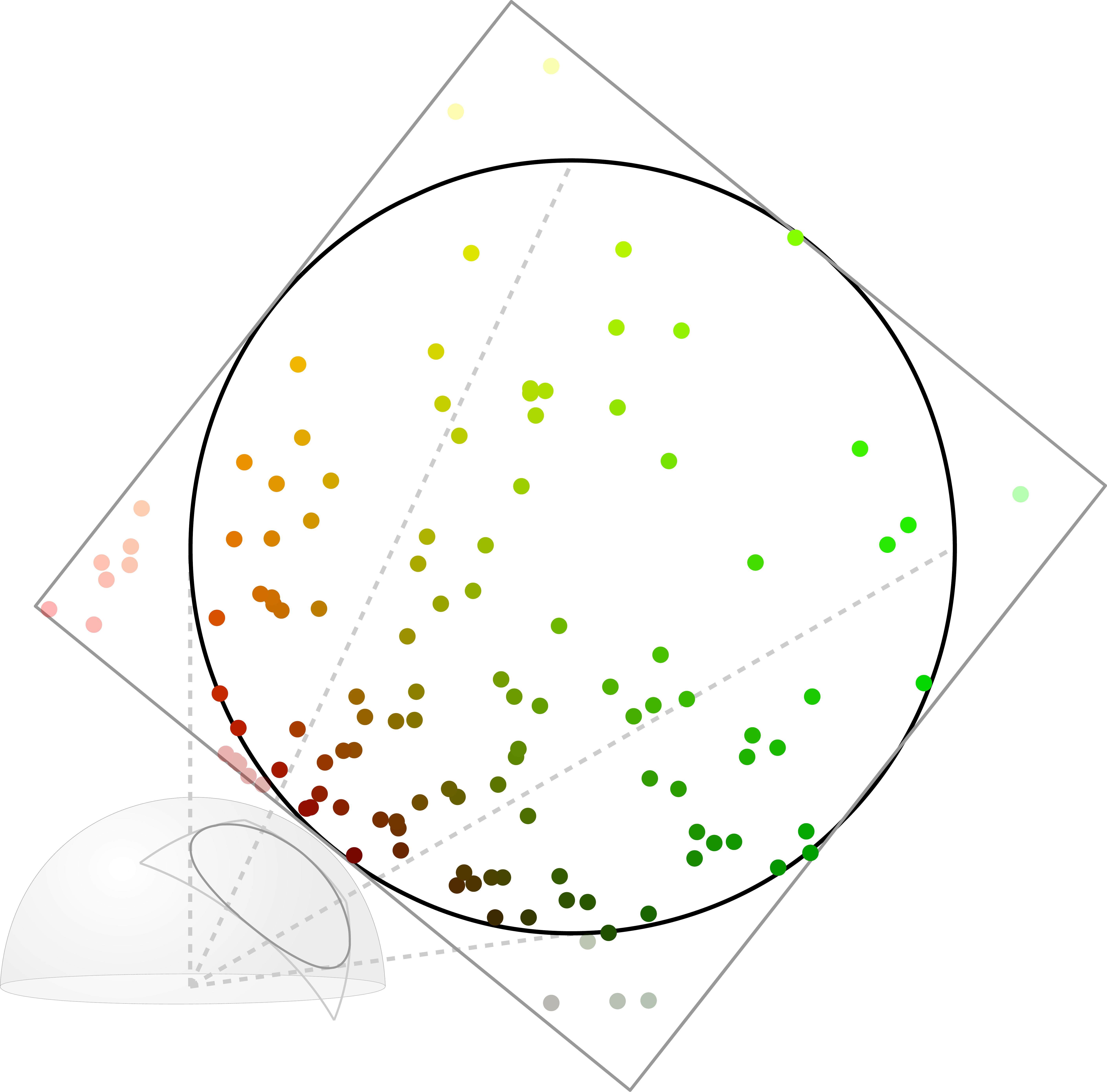
		\hspace{0.18cm}
		\def\svgwidth{\fithlinewidth}
		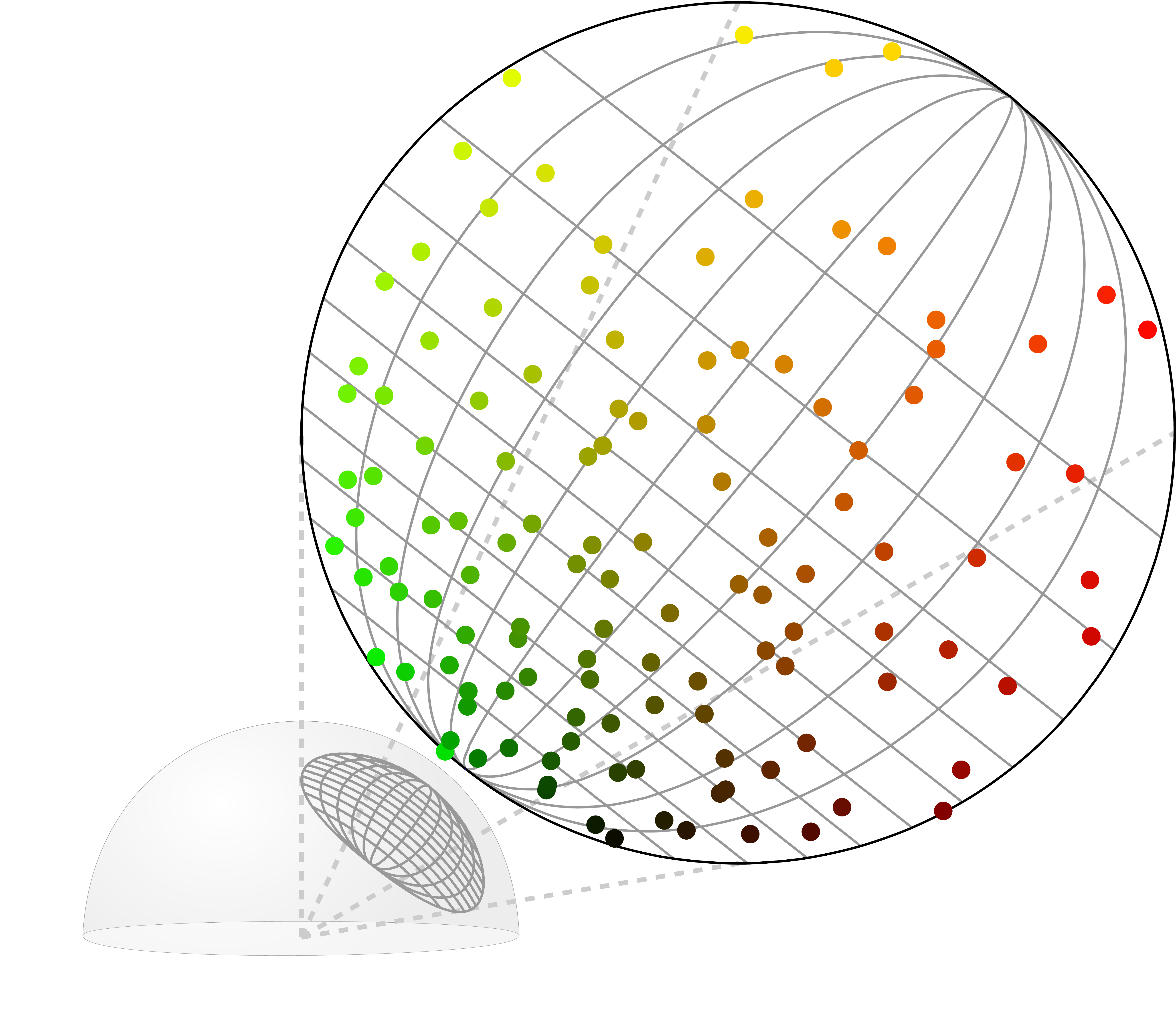
		\hspace{0.22cm}
		\def\svgwidth{\fithlinewidth}
		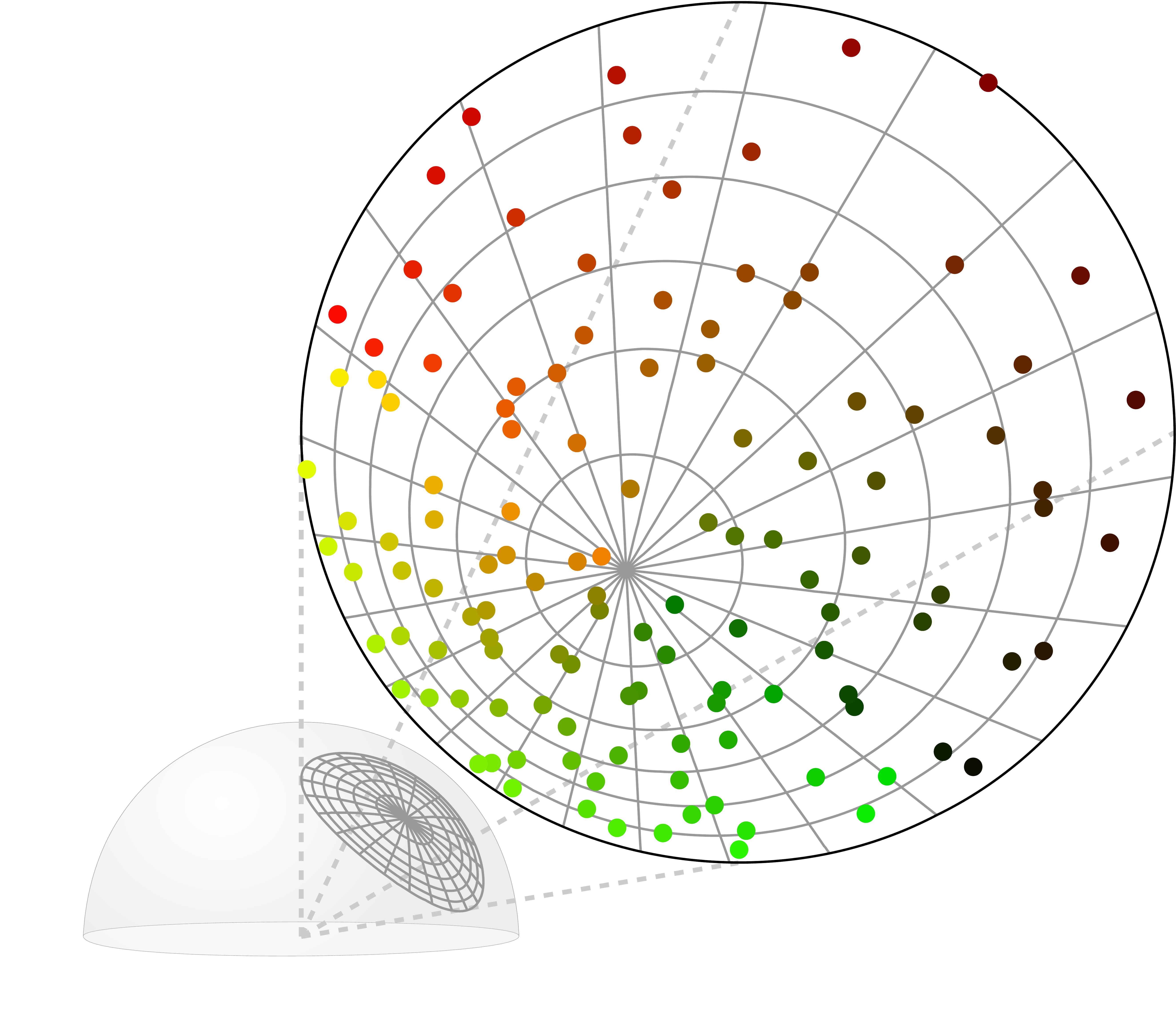
		\hspace{0.22cm}
		\def\svgwidth{\fithlinewidth}
		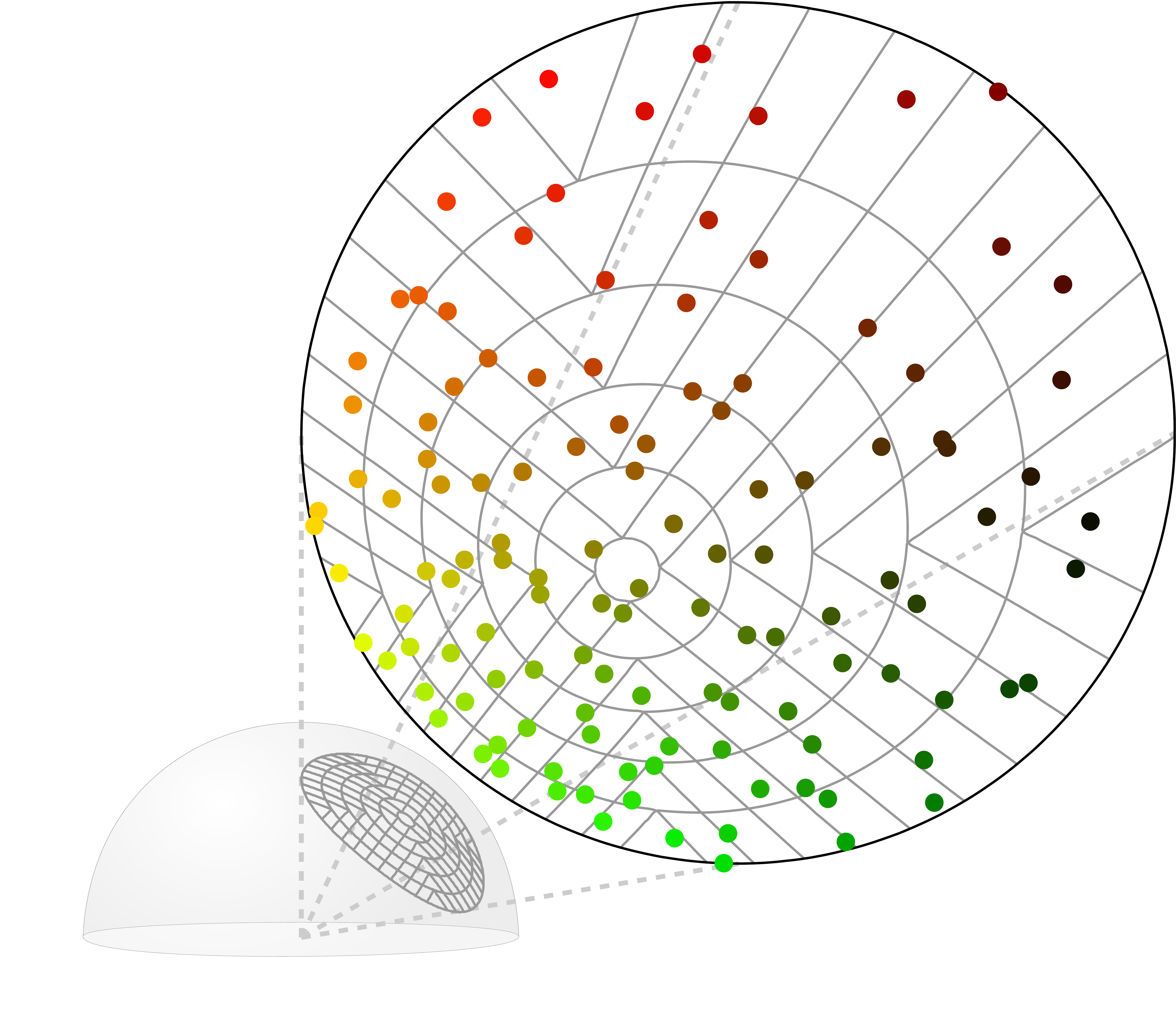 \\
		\begin{subfigure}{\fithlinewidth}
		\def\svgwidth{\linewidth}
		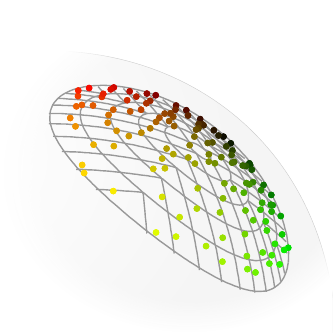
		\vspace{-1.8em}
		\caption{Area sampling~\cite{Shirley:1997:Disks}}
		\label{fig:disk_area}
		\end{subfigure}
		\hspace{0.22cm}
		\begin{subfigure}{\fithlinewidth}
		\def\svgwidth{\linewidth}
		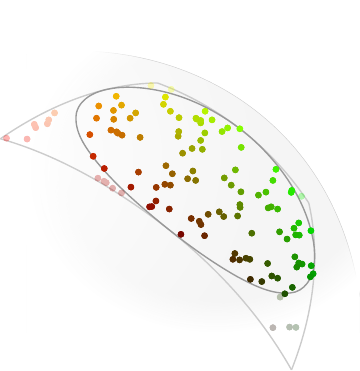
		\vspace{-1.8em}
		\caption{Gamito~\cite{Gamito:2016:Disks}}
		\label{fig:disk_gamito}
		\end{subfigure}
		\hspace{0.22cm}
		\begin{subfigure}{\fithlinewidth}
		\def\svgwidth{\linewidth}
		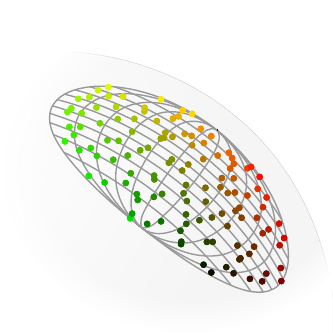
		\vspace{-1.8em}
		\caption{Our parallel mapping}
		\label{fig:disk_urena}
		\end{subfigure}
		\hspace{0.22cm}
		\begin{subfigure}{\fithlinewidth}
		\def\svgwidth{\linewidth}
		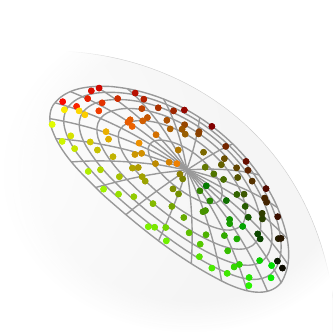
		\vspace{-1.8em}
		\caption{Our radial mapping}
		\label{fig:disk_ibon}
		\end{subfigure}
		\hspace{0.22cm}
		\begin{subfigure}{\fithlinewidth}
		\def\svgwidth{\linewidth}
		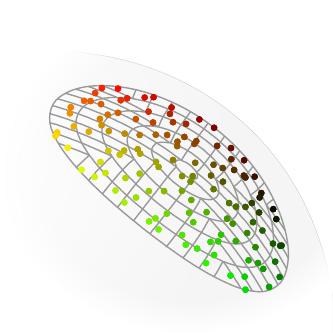
		\vspace{-1.8em}
		\caption{Our ld-radial mapping}
		\label{fig:disk_alan}
		\end{subfigure}
		\vspace{-0.5em}
	\caption{A stratified unit-square sample pattern transformed onto the surface of a disk using existing techniques and our proposed maps (with solid angle projections on the bottom row). The points are colored according to their canonical $[0,1]^2$ coordinates to illustrate the continuity of the maps. Gamito's rejection sampling does not allow for direct stratification, so we show the candidate low-discrepancy pattern for that case.}
	\label{fig:disks}
\end{figure*}
%


\section{Problem Statement and Previous Work}
\label{sec:problem}

Our goal is to compute the radiance
$\Ls$ scattered at a point $\px$ in direction $\wo$ due to irradiance from a disk-shaped luminaire $D$.
This can be written as an integral over the solid angle $\lightSolidAngle$ subtended by the luminaire:
\begin{equation}
   \label{eq:rendering}
   \Ls(\px,\wo) =  \int_{\lightSolidAngle} \contrib(\px, \py_{\px\ww}, \wo, \ww) \,\text{d}\mu(\ww),
\end{equation}
where $\py_{\px\ww}$ is the first visible point from $\px$ in direction $\ww$, $\mu$ is the solid angle measure, and the contribution function $\contrib$ is
\begin{equation*}
   \contrib(\px, \py, \wo, \ww) \! = \!
   \begin{cases}
      \Le(\py, \!\shortminus\ww) \brdf(\px,\wo,\ww) |\ww \cdot \normal{n}_\px|, & \!\!\!\!\text{if $\px$ is on a surface,}\\
      \Le(\py, \!\shortminus\ww) \bsdf{\px,\wo,\ww} T(\px,\py), & \!\!\!\!\text{if $\px$ is in a medium,}
   \end{cases}
\end{equation*}
with $\brdf$, $\normal{n}_\px$, and $\rho$ being respectively the BSDF, surface normal, and medium phase function (times the scattering coefficient) at $\px$. $\Le(\py, \shortminus\ww)$ is the luminaire emission radiance at $\py$ in direction $-\ww$ and $T(\px,\py)$ is the medium transmitance between $\px$ and $\py$.

\paragraph*{Solid angle sampling.}
Monte Carlo estimation of Equation~\eqref{eq:rendering} using $N$ randomly sampled directions $\ww_i$ has the following form:
\begin{equation}
  \label{eq:rendering_estimator}
  \Ls(\px,\wo) \approx \frac{1}{N} \sum_{i=1}^N \frac{\contrib(\px, \py_{\px\ww_i}, \wo, \ww_i)}{p(\ww_i)},
\end{equation}
where $p(\ww)$ is the pdf for sampling $\ww$. The choice of sampling density $p$ is important,
since a lower variation of $f/p$ makes the estimator more efficient~\cite{Shirley:1996:Direct}.
For disk lights the traditional choice is uniform density over the luminaire surface $D$. This \emph{area sampling} technique is easy to implement and its resulting solid angle pdf is $p(\ww) = \|\px - \py_{\px\ww}\| / \left(A(D) \, |\ww \cdot \normal{n}_{\py_{\px\ww}}|\right)$, where $A(D)$ is the area of $D$. This pdf can lead to very high variance in the radiance estimator~\eqref{eq:rendering_estimator}, especially when the point $\px$ is close to the luminaire. Our goal in this paper is to devise uniform \emph{solid angle sampling} techniques that generate directions $\ww$ with constant density $p(\ww) = 1 / |\lightSolidAngle|$, yielding estimators with significantly lower variance than uniform area sampling.

\paragraph*{Area-preserving mapping.}
Sample stratification can greatly improve the efficiency of Monte Carlo estimators~\cite{Shirley:1991:Discrepancy,Subr:2013:FASS,Pilleboue:2015:Variance}.
Most existing stratification techniques generate samples in the
canonical unit square $[0,1]^2$, however our goal is to sample directions inside the solid
angle $\lightSolidAngle$. Therefore, in order to take advantage of these
techniques, we need to find a mapping $M$ from $[0,1]^2$ to $\lightSolidAngle$ such that for any
two regions $R_1, R_2 \subseteq [0,1]^2$:
\begin{equation*}
    \frac{A(R_1)}{A(R_2)} = \frac{\mu(M(R_1))}{\mu(M(R_2))},
\end{equation*}
where $A$ is the area measure, and $\mu$ is the solid angle measure as in Equation~\eqref{eq:rendering}. We call such maps \emph{area-preserving maps}. This key
property makes it possible to generate stratified samples in $\lightSolidAngle$,
because stratification is far more easily achieved in $[0,1]^2$.

Area-preserving solid angle maps have been developed for
triangles~\shortcite{Arvo:1995:Triangles} and rectangles~\shortcite{Urena:2013:Quads}.
For sampling the solid angles of disks, Gamito~\shortcite{Gamito:2016:Disks} proposed to use
a rectangle map~\shortcite{Urena:2013:Quads} followed by rejection sampling. This technique
cannot be used with fixed-size canonical point sets, and needs a
low-discrepancy sequence capable of progressively generating stratified sample candidates.
The rejection sampling also makes it very difficult to achieve good high-dimensional
stratification in the presence of other distributed effects, e.g.\ volumetric
scattering, where the coordination of the sample patterns of different effects is desired.
In this paper we focus on area-preserving maps for disks that do not require
rejection sampling and work with any canonical sample pattern. \Fref{fig:disks} compares
our proposed maps against existing techniques.

For surface scattering points $\px$, an even better
strategy is to importance sample the term $|\ww \cdot \normal{n}_\px|$ in the
contribution $f$. Such uniform sampling of the \emph{projected solid angle} has been described by
Arvo~\shortcite{Arvo:2001:Manifolds} for polygonal lights. Extending our approach
to projected solid angle sampling is an interesting avenue for future work.


\begin{figure*}[t]
\centering
	\def\svgwidth{.9\textwidth}
	\fontsize{0.26cm}{1em}
	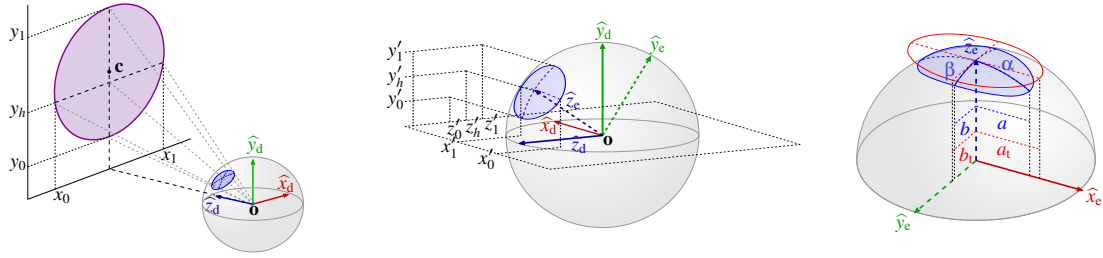
        \vspace{-0.5em}
	\caption{
	{\itshape Left:} The disk's local reference system $\refSystem{d}=(\axis{x}[d],\axis{y}[d],\axis{z}[d])$ and the local coordinates required to characterize its solid angle projection.
	{\itshape Center:} Projections of the relevant coordinates onto the unit sphere, defining the spherical ellipse and its local reference system $\refSystem{e}=(\axis{x}[e],\axis{y}[e],\axis{z}[e])$, where $\axis{x}[e] \equiv \axis{x}[d]$.
	{\itshape Right:} The spherical ellipse is defined by its semi-arcs $\alpha$ and $\beta$ or, equivalently, by its semi-axes $a$ and $b$ in $\refSystem{e}$. The tangent ellipse (in red), which lies on a plane tangent to the sphere at $\axis{z}[e]$ (i.e.\ the spherical ellipse center), is defined by its semi-axes $\atang$ and $\btang$.
	}
	\label{fig:projection}
\end{figure*}

\vspace{0.2em}

\section{Solid Angle Sampling of an Oriented Disk}
\label{sec:desarrollo}

We base our sampling techniques on the key observation that the projected area of any ellipse, including a disk, forms a spherical ellipse on the unit sphere around the shading point (\Fref{fig:projection}). Thus, in order to sample the solid angle subtended at point $\point{o}$ by an oriented disk with center $\point{c}$, normal $\normal{n}$, and radius $r$, we will uniformly sample a point $\point{q}$ on the spherical ellipse and then backproject it to the disk.

\paragraph*{Spherical ellipse.} To compute the subtended spherical ellipse, we first define a local reference frame for the disk $\refSystem{d}=(\axis{x}[d],\axis{y}[d],\axis{z}[d])$:
\begin{align}
\axis{z}[d] &= -\normal{n}, & \axis{y}[d] &= \axis{z}[d] \times \frac{\point{c} - \point{o}}{\|\point{c} - \point{o}\|}, & \axis{x}[d] &= \axis{y}[d] \times \axis{z}[d].
\end{align}
We then take the boundary disk coordinates $y_0$ and $y_1$ w.r.t.\ the $\axis{y}[d]$ axis and project them onto the sphere (\Fref{fig:projection}, left). From the coordinates $y'_0$, $y'_1$, $z'_0$, $z'_1$ of these projections (\Fref{fig:projection}, middle) we can compute the spherical ellipse center: it is the result $\axis{z}[e]$ of normalizing the vector $(0, y'_h,z'_h)$, where $y'_h = (y'_0 + y'_1)/2$ and $z'_h = (z'_0 + z'_1)/2$. (Note that $\axis{z}[e]$ in general does not coincide with the spherical projection of the disk center $\point{c}$.) Reprojecting $\axis{z}[e]$ onto the disk (\Fref{fig:projection}, left), the obtained $y_h$ coordinate defines a chord $\overline{x_0x_1}$ parallel to $\axis{x}[d]$. The chord endpoint projections onto the sphere, with $\axis{x}[d]$-coordinates $x'_0$ and $x'_1$, allow us to compute the lengths of the ellipse's semi-axes, $a$ and $b$, and semi-arcs, $\alpha$ and $\beta$ (\Fref{fig:projection}, right):
\begin{align}
	a &= x_1', &
	b &= \frac{1}{2}\sqrt{(y_1' - y_0')^2 + (z_1' - z_0')^2},\\
	\alpha &= \sin^{-1}{a}, &
	\beta  &= \sin^{-1}{b}.
\end{align}
Finally, from $\alpha$ and $\beta$ we can compute the semi-axes $\atang=\tan\alpha$ and $\btang=\tan\beta$ of the ellipse tangent to the sphere at $\axis{z}[e]$ (\Fref{fig:projection}, right).

In the following,
we use both the spherical and the tangent ellipses to derive two
different mappings for uniformly sampling points $\point{q}$ on the
spherical ellipse which we then map to the surface of the disk.
These mappings operate in a coordinate system $\refSystem{e}=(\axis{x}[e],\axis{y}[e],\axis{z}[e])$, where $\axis{x}[e] \equiv \axis{x}[d]$ and $\axis{y}[e] = \axis{z}[e] \times \axis{x}[e]$, shown in Figures \ref{fig:projection} and \ref{fig:disk_area}.

\begin{figure}[b!]
	\vspace{-0.2em}
	\fontsize{0.26cm}{1em}
	\def\svgwidth{\linewidth}
	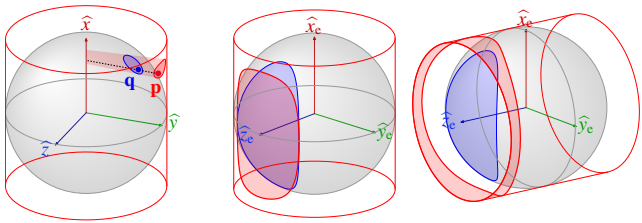
	\vspace{-1.8em}
	\caption{
		{\itshape Left}: Any region (blue) on the unit sphere can be
		radially projected to another region (red) on a cylinder
		aligned with any axis (here $\axis{y}$). Any point $\point{q}$ on the sphere  can be expressed in cylindrical coordinates
		as $(\phi,h,r)$ (azimuth angle, altitude, distance from center). This point can be
		mapped to a point $\point{p}$ on the unit cylinder with coordinates $(\phi,h)$. The mapping
		preserves the areas of both finite and differential regions. Thus, to obtain
		a point on the spherical region, we can sample inside the cylindrical region and
		project back onto the sphere.
		{\itshape Center}: A spherical ellipse (blue), with center on
		the $\axis{z}[e]$ axis, projected onto a cylinder aligned with the $\axis{y}[e]$ axis.
		{\itshape Right}: The same spherical ellipse projected onto a $\axis{z}[e]$-axis
		aligned cylinder. In this case, the projected region (red) has a ring-like shape.
	}
	\label{fig:disk_area}
	\vspace{-0.3em}
\end{figure}

\subsection{Area-preserving mappings}

Our new mappings are based on a
generalization of the so-called \textit{Archimedes Hat-Box theorem}. It
states that the area of a region on the sphere between two parallels is
equal to the area of that region's projection onto a perpendicularly aligned bounding cylinder.
This area-preserving
property also holds for arbitrarily shaped regions on the sphere
(Figure~\ref{fig:disk_area}, left). The latter property was used by Tobler et al.
\cite{Tobler1998} to define compact metallic BRDFs (they provide a
demonstration). It was also used (although not explicitly stated) by Arvo
\cite{Arvo:1995:Triangles} and Ureña et al. \cite{Urena:2013:Quads} to
define area-preserving parameterizations for spherical triangles and rectangles,
respectively.

We use this cylindrical projection property to derive our area-preserving mappings for a spherical ellipse centered on the $\axis{z}[e]$ axis. The
ellipse can be radially projected onto a cylinder, obtaining a
{\itshape cylindrical ellipse}. Two different unit-radius cylinders can be used.
One is aligned with the $\axis{y}[e]$ axis (Figure~\ref{fig:disk_area}, center), which
we call a \emph{parallel map}. The second one is aligned with the
$\axis{z}[e]$ axis (Figure~\ref{fig:disk_area}, right), which we call a \emph{radial map}.
We also propose a variant of the radial map
that uses Shirley's low-distortion map \cite{Shirley:1997:Disks}, which we call a
\emph{low-distortion radial map}, or \emph{ld-radial map}.

\paragraph*{Maps overview.}
The basic idea behind our maps is to first select a point $\point{p}$ on the
cylindrical ellipse as a function of a canonical unit-square point $(\randVar[1],\randVar[2]) \in [0,1]^2$. We then project $\point{p}$ back onto the
sphere perpendicularly to the cylinder axis (see Figure \ref{fig:disk_area},
left) to get the point $\point{q}$. Let $(\phi,h)$ be the
cylindrical coordinates of $\point{p}$. We first obtain the azimuth angle $\phi$ by finding the lateral slice on the
cylindrical ellipse whose solid angle is $\randVar[1]\lightSolidAngle$ (\Fref{fig:urena_integration}). With $\phi$ fixed, the altitude $h$ is computed as a simple linear interpolation using $\randVar[2]$ along the lateral line segment that is the intersection between the lateral plane at angle $\phi$ and the cylindrical
ellipse (green line segment in Figure \ref{fig:urena_integration}). The sampling of $\phi$ involves numerical inversion of incomplete elliptic integrals, as we show next.

\subsection{Parallel Mapping}
\label{sec:urena_mapping}

\begin{figure}[t!]
	\centering
  \def\svgwidth{37\linewidth/100}
	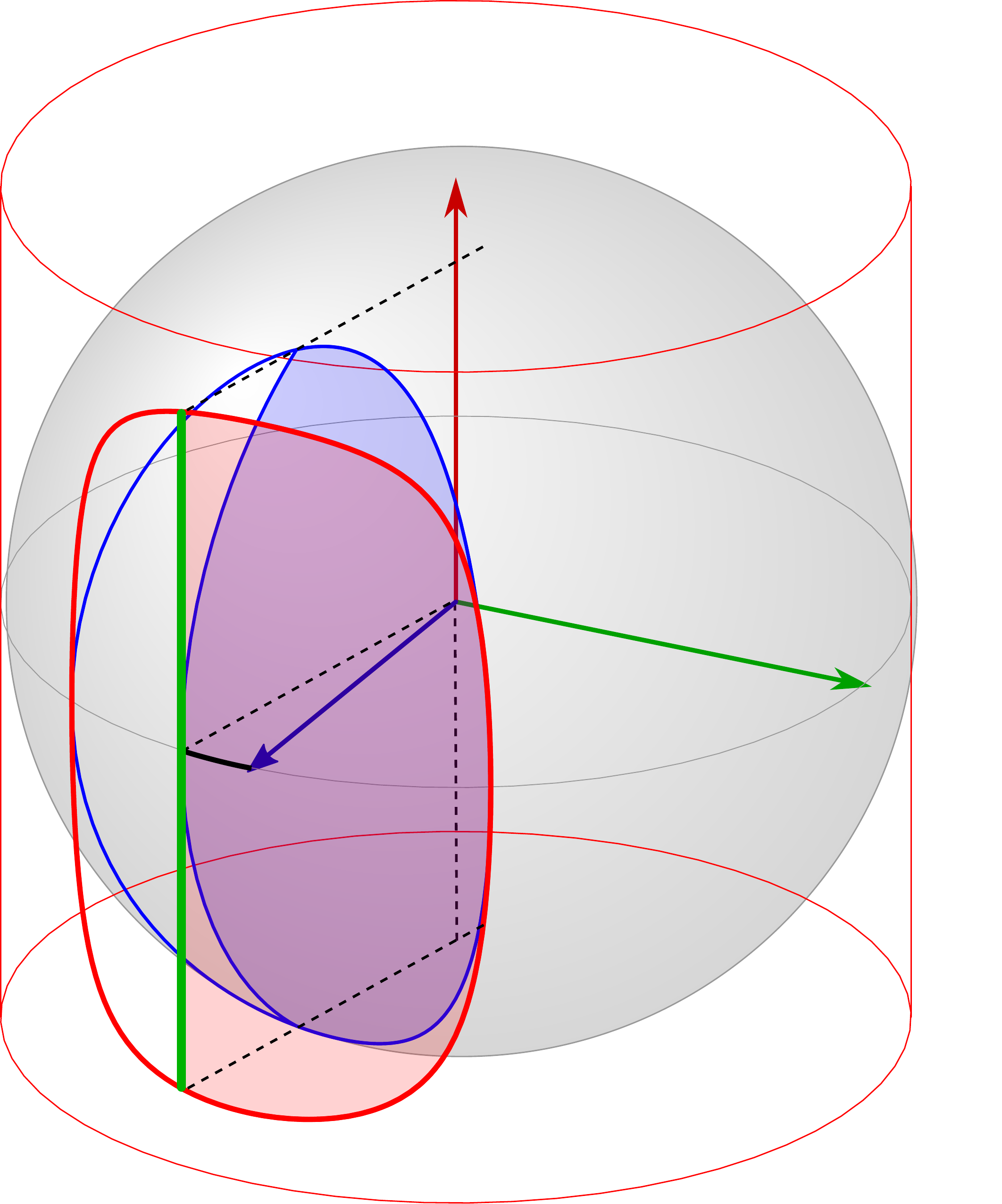
	$~~~~~~~$
  \def\svgwidth{53\linewidth/100}
	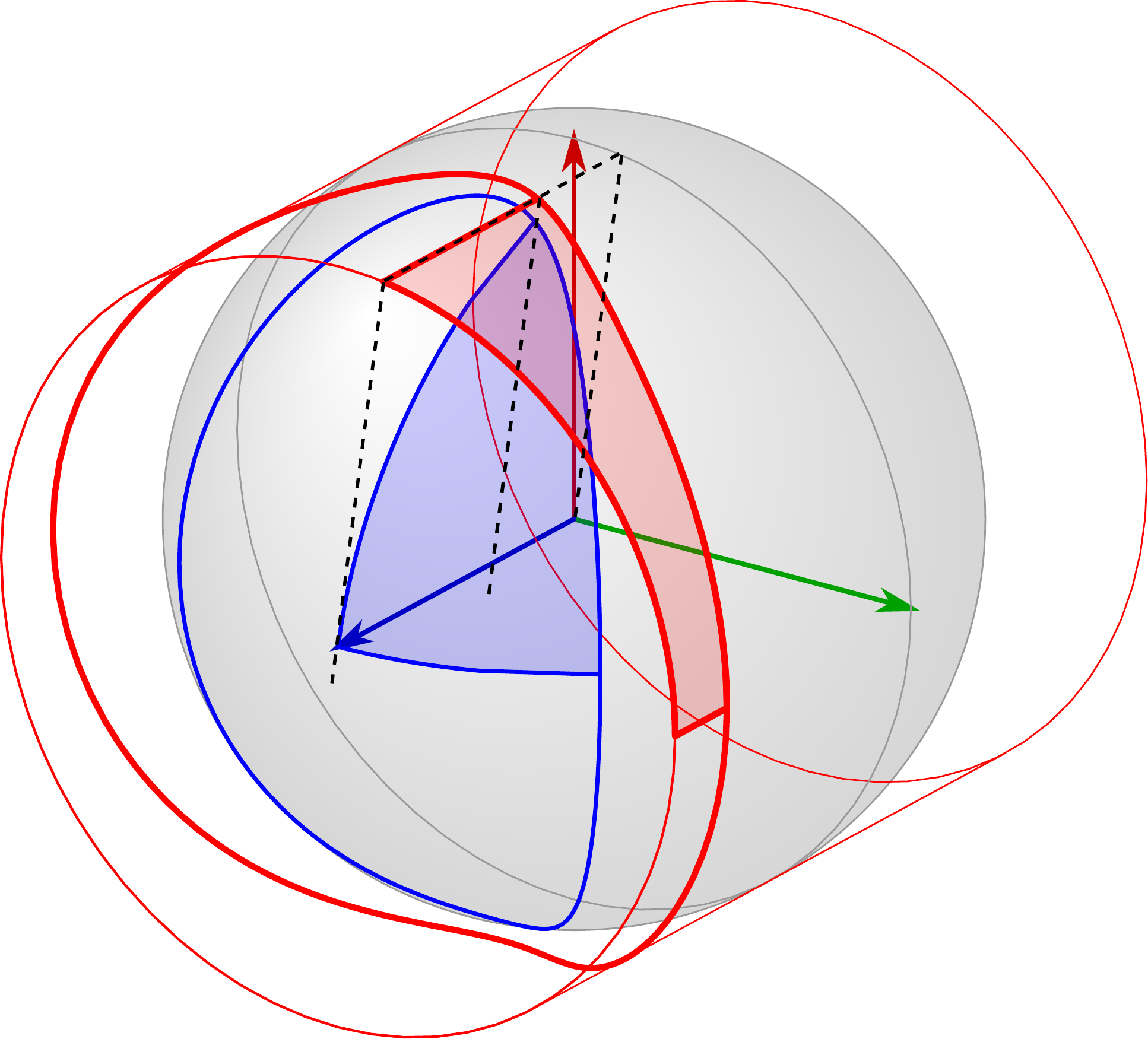
	\vspace{-0.3em}
	\caption{%
	Illustration of our parallel (left) and radial (right) maps.
	Given a canonical sample $(\randVar[1], \randVar[2]) \in [0;1]^2$, we first find the azimuth angle $\anglePolar$, respectively $\angleRadial$, that cuts a region on the cylindrical ellipse with area $\epsilon_1\lightSolidAngle$~(in red, determined by the green line segment).
	A sample point on the cylinder is then obtained by linearly interpolating the green segment endpoints using $\randVar[2]$.
	For the parallel map, the endpoint cylindrical coordinates are $(\anglePolar,-\hPolar)$ and $(\anglePolar,\hPolar)$, with $\anglePolar \in [-\beta,\beta]$. For the radial map, these coordinates are $(\angleRadial,\hRadial)$ and $(\angleRadial,1)$, with $\angleRadial \in [0,2\pi]$ (we use $\angleRadial \in [0,\pi/2]$ in each quadrant).~~~~} 
	\label{fig:urena_integration}
	\vspace{-0.5em}
\end{figure}

Our parallel mapping, whose cylinder axis is aligned with $\axis{x}[e]$, operates by considering a portion (sector) of the cylindrical ellipse -- the red-shaded region in Figure~\ref{fig:urena_integration}, left. This region is determined by the green line segment, whose endpoints have cylindrical coordinates $(\anglePolar,\hPolar)$ and $(\anglePolar,-\hPolar)$. The angle $\anglePolar$ goes from
$-\beta$ to $\beta$, since the spherical ellipse is centered on the $\axis{z}[e]$ axis.

Due to the Hat-Box theorem, the differential solid angle covered
by the green segment is equal to its length, $2\hPolar$, which is in fact a function of $\anglePolar$. Thus, the solid angle subtended by the red region onto the spherical ellipse (the blue region in Figure~\ref{fig:urena_integration}, left) can be written as the integral of the segment length:
\begin{equation}
	\solidAnglePolar(\anglePolar) = \int_{-\beta}^{\anglePolar} 2\hPolar(\anglePolar') \diff \anglePolar',
	\label{eq:integral_omega_urena}
\end{equation}
where the full solid angle of the spherical ellipse is $\lightSolidAngle=\solidAnglePolar(\beta)$.
Due to symmetry, for any angle $\anglePolar \!\in\! [0,\beta]$ it holds $\hPolar(\!-\anglePolar) = \hPolar(\anglePolar)$. We use this to express $\solidAnglePolar(\anglePolar)$ as a sum of integrals $\solidAnglePolarPos(\anglePolar)$ over positive angles:
\begin{equation}
	\solidAnglePolar(\anglePolar) =
		\begin{cases}
			\solidAnglePolarPos(\beta) + \solidAnglePolarPos(\anglePolar)   &: \anglePolar \ge 0 \\
			\solidAnglePolarPos(\beta) - \solidAnglePolarPos(-\anglePolar) &: \anglePolar < 0
		\end{cases},
	\label{eq:delta_pos_int}
\end{equation}
where
\begin{equation}
	\solidAnglePolarPos(\anglePolar) = \int_{0}^{\anglePolar} 2\hPolar(\anglePolar') \diff \anglePolar'.
	\label{eq:delta_pos_int1}
\end{equation}
In Appendix~\ref{sec:h1deriv} we derive an expression for $\hPolar(\anglePolar)$:
\begin{equation}
	\hPolar(\anglePolar) = \ctang \sqrt{\dfrac{1 - (p + 1) \sin^2{\anglePolar}}
									 {1 - (m\,p + 1) \sin^2{\anglePolar}}},
	\label{eq:delta_sin}
\end{equation}
where
\begin{align}
	p & = \dfrac{1}{\btang^2},  &
	m & = \dfrac{\atang^2 - \btang^2}{\atang^2 + 1} , &
	\ctang & = \dfrac{\atang}{\sqrt{1 + \atang^2}}.
	\label{eq:pmc}
\end{align}
Substituting \Fref{eq:delta_sin} into \Fref{eq:delta_pos_int1} and simplifying, we get:\!
\begin{equation}
	\solidAnglePolarPos(\anglePolar) =
	\frac{2\ctang}{\btang}
	\Big[
	   (1 - n) \thirdKind{n}[\varphi_\text{p}]{m}
		 -
     \firstKind[\varphi_\text{p}]{m}
	\Big],
	\label{eq:urena_int_real}
\end{equation}
where $\firstKind[\varphi]{m}$ and $\thirdKind{n}[\varphi]{m}$ are Legendre incomplete elliptic integrals of respectively the first and third kind, $m \in [0,1)$, and
\begin{equation}
	\varphi_\text{p} =  \sin^{-1}{\left( \dfrac{\tan{\anglePolar}}{\btang} \right)},
	~~~~~~~~~~~
	n = -\btang^2.
\end{equation}
Unfortunately, no closed-form expressions are known for $\firstKind[\varphi]{m}$ and $\thirdKind{n}[\varphi]{m}$, so \Fref{eq:urena_int_real} must be evaluated numerically.

\paragraph*{Sampling.}
With the fractional spherical ellipse area $\solidAnglePolar$ characterized, we can map a point on the unit square $(\randVar[1], \randVar[2]) \in [0,1]^2$ to a point on the spherical ellipse $\point{q}$.
We first need to find the angle $\anglePolar$ that satisfies $\solidAnglePolar(\anglePolar)=\randVar[1]\lightSolidAngle$, for which we need to evaluate the inverse function $\solidAnglePolar^{-1}$. This function has no analytical form, so we resort to numerically finding the roots of the equation
\begin{equation}
	\solidAnglePolar(\anglePolar) - \randVar[1] \lightSolidAngle = 0.
	\label{eq:urena_root_finding}
\end{equation}
Having sampled $\anglePolar$, we get the point $\point{p}$ on the cylindrical ellipse by first computing $\hPolar(\anglePolar)$ using \Fref{eq:delta_sin} and then linearly interpolating the altitude coordinate between $-\hPolar(\anglePolar)$ and $\hPolar(\anglePolar)$ using $\randVar[2]$:
\vspace*{-1.2em}
\begin{equation}
	\point{p} = (\anglePolar,(2\randVar[2]-1)\hPolar(\anglePolar)) = (\anglePolar, h).
\end{equation}
Finally, the corresponding point $\point{q}$ on the ellipse is obtained by radially projecting $\point{p}$ onto the sphere (see \Fref{fig:disk_area}, left):
\begin{equation}
	\point{q} =
	  \left(
		  h, \,
			\sqrt{1 - h^2} \sin{\anglePolar}, \,
		  \sqrt{1 - h^2} \cos{\anglePolar}
		\right).
	\label{eq:point_qpolar}
\end{equation}
\Fref{fig:disk_urena} shows the resulting map.

\subsection{Radial Mapping}
\label{sec:ibon_mapping}



The parallel mapping presented in \Fref{sec:urena_mapping} involves two elliptic integrals and introduces noticeable distortions (see the converging lines in \Fref{fig:disk_urena}), which can increase discrepancy and ruin any blue noise properties present in the input unit-square sample distribution. In this section we present an alternative radial mapping that uses a single elliptical integral and also exhibits less distortion. It is based on the analysis of the spherical ellipse topology by Booth~\shortcite{Booth:1844:SphericalEllipse}.

We will exploit the fact that the four quadrants of the spherical ellipse are radially symmetric (see \Fref{fig:projection}, right), so its total area can be expressed as $\lightSolidAngle = 4 \, \solidAngleRadial$, with $\solidAngleRadial$ being the area of each quadrant. Within a quandant, the azimuth angle is $\angleRadial \in [0, \pi/2]$.

\begin{figure}[t!]
	\centering
	\def\svgwidth{\linewidth}
	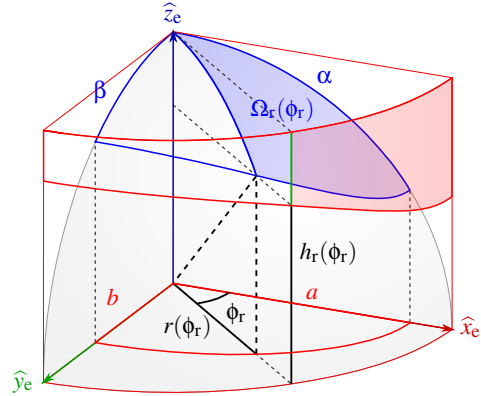
	\caption{Illustration of our polar mapping. Projected perpendicularly to its axis $\axis{z}[e]$ onto the $\axis{x}[e]\axis{y}[e]$ plane, the spherical ellipse forms a planar ellipse with semi-axes $a = \sin{\alpha}$ and $b = \sin{\beta}$.
	We use the Pythagorean theorem to express the altitude $\hRadial(\angleRadial)$ of the cylindrical projection of the spherical ellipse's curve in terms of $a$, $b$, and $\angleRadial$.}
	\label{fig:ellipse_quadrant}
\end{figure}

We now consider a bounding cylinder aligned with the $\axis{z}[e]$ axis (\Fref{fig:urena_integration}, right). Specifically, we are interested in the lateral region (in red in \Fref{fig:ellipse_quadrant}) that is the radial projection of a fraction of the spherical quadrant. This region is determined by the position of the green line segment whose endpoints have cylindrical coordinates $(\angleRadial, \hRadial)$ and $(\angleRadial, 1)$. The segment length is $1 - \hRadial$, which is a function of $\angleRadial$. Similarly to \Fref{eq:integral_omega_urena}, we use the Hat-Box theorem to express the fractional quadrant area as the integral of this length:
\begin{equation}
	\solidAngleRadial(\angleRadial) =
	\int_0^{\angleRadial}
		\left[ 1 - \hRadial(\angleRadial') \right] \diff \angleRadial' =
	\angleRadial - \int_0^{\angleRadial}
		\hRadial(\angleRadial') \diff \angleRadial'.
	\label{eq:integral_delta_booth}
\end{equation}
Using the Pythagorean theorem, we express $\hRadial(\angleRadial)$ as (see \Fref{fig:ellipse_quadrant})\!\!
\begin{equation}
	\hRadial(\angleRadial) = \sqrt{1 - r^2(\angleRadial)},
	\label{eq:height_polar}
\end{equation}
where $r(\angleRadial)$ is the (planar) elliptical radius of the spherical ellipse with the following expression, which we derive in Appendix~\ref{sec:deriv_ellipse_radius}:
\begin{equation}
	r(\angleRadial) = \frac{a b}{\sqrt{a^2 \sin^2{\angleRadial} + b^2 \cos^2{\angleRadial}}}.
	\label{eq:ellipse radius}
\end{equation}

Plugging Equations~\eqref{eq:height_polar} and \eqref{eq:ellipse radius} back into \Fref{eq:integral_delta_booth}, and using Booth's derivations \shortcite{Booth:1844:SphericalEllipse}, we can now express the fractional quadrant area $\solidAngleRadial(\angleRadial)$ using Legendre's incomplete elliptic integral of the third kind $\thirdKind{n}[\varphi]{m}$, so it becomes\!
\begin{equation}
	\solidAngleRadial(\angleRadial) =
		\angleRadial - \frac{b (1 - a^2)}{a \sqrt{1 - b^2}} \,\thirdKind{n}[\varphi_\text{r}]{m},
	\label{eq:elliptic_integral}
\end{equation}
where
\begin{align}
	\!\!\!\!
	\varphi_\text{r} \!&=\! \tan^{\!-1}\!\!\left( \frac{\atang}{\btang} \tan{\angleRadial} \!\right)\!, \, &
	n  \! &= \! \frac{a^2 \! - b^2}{a^2 (1 - b^2)}, \, &
	m  \! &= \! \frac{a^2 \! - b^2}{1 - b^2}. \,
\end{align}
Above, $\varphi_\text{r}$ is the parametric angle of the tangent ellipse, and $n$ and $m$ are the elliptic characteristic and module that characterize the elliptic integral~\cite{Booth:1852:EllipticIntegrals}. 

Unfortunately, as with \Fref{eq:urena_int_real}, no closed-form expression is known for the general-case incomplete elliptic integral of the third kind, so we need to evaluate \Fref{eq:elliptic_integral} numerically.

\paragraph*{Direct radial mapping.}
Having an expression for the fractional spherical ellipse $\solidAngleRadial$, we can map a unit-square point $(\randVar[1], \randVar[2]) \in [0,1]^2$ to a point $\point{q}$ on the spherical ellipse. Below we only consider sampling the first ellipse quadrant (shown in \Fref{fig:ellipse_quadrant}); the entire ellipse can be covered by flipping the $\axis{x}[e]$- and $\axis{y}[e]$-coordinates of $\point{q}$.

First, we need to find the azimuth angle $\angleRadial \in [0,\slfrac{\pi}{2}]$ satisfying
\begin{equation}
	\solidAngleRadial(\angleRadial) - \randVar[1] \, \solidAngleRadial(\slfrac{\pi}{2}) = 0.
	\label{eq:ibon_root_finding}
\end{equation}
Since we do not have a method to analytically invert $\solidAngleRadial(\angleRadial)$, we compute $\angleRadial$ by numerically finding the roots of the above equation.

Having sampled $\angleRadial$, we obtain point $\point{p}$ on the cylindrical ellipse by first computing $\hRadial(\angleRadial)$ using \Fref{eq:height_polar} and then linearly interpolating the altitude coordinate between $\hRadial(\angleRadial)$ and $1$ using $\randVar[2]$:
\begin{equation}
	\point{p} = \big( \angleRadial,\, (1 - \randVar[2]) \hRadial(\angleRadial) + \randVar[2] \big) = (\angleRadial, h).
	\label{eq:point_ppolar}
\end{equation}
We find the corresponding point $\point{q}$ on the ellipse by projecting $\point{p} = (\angleRadial, h)$ using \Fref{eq:point_qpolar} with swapped $\axis{x}[e]$- and $\axis{z}[e]$-coordinates. \Fref{fig:disk_ibon} shows the resulting map.

\paragraph*{Low-distortion radial mapping.}
As seen in \Fref{fig:disk_ibon}, the direct mapping from above resembles the classical planar Cartesian-to-polar mapping. As such, it also exhibits the same distortion -- the lines converging at the ellipse center, which does not preserve relative distances between samples and damages their stratification. In the planar case, the mapping of Shirley and Chiu~\shortcite{Shirley:1997:Disks} rectifies this distortion by warping concentric squares into concentric disks. To achieve the analogous mapping on the spherical ellipse, we first warp our input unit-square samples $(\randVar[1], \randVar[2])$ to the unit disk using Shirley and Chiu's concentric mapping. We then move back to the unit square using the following \emph{inverse polar mapping}:
\begin{align}
	u &=
		\begin{cases}
			2 \, \dfrac{\theta}{{\pi}} &: \
				\theta \in [0, \frac{\pi}{2}) \\
			1 - 2 \, \dfrac{\theta -
					{\pi}/{2}}{{\pi}} &: \
				\theta \in [\frac{\pi}{2}, \pi) \\
			2\,\dfrac{\theta -
					\pi}{{\pi}} &: \
				\theta \in [\pi, \frac{3\pi}{2}) \\
			1 - 2\,\dfrac{\theta - {3/2\pi}}
					  {{\pi}} &: \
				\theta \in [\frac{3\pi}{2}, 2\pi) \\
		\end{cases}, &
	v &= r^2.
\end{align}
The result of this detour is a unit-square point set that, when warped using the classical (forward) planar polar mapping, gives Shirley and Chiu's low-distortion concentric disk distribution. We, instead, feed this unit-square set to our direct radial mapping to get a concentric-like distribution on the spherical ellipse, which is shown in \Fref{fig:disk_alan}.

\section{Implementation}

We have implemented our maps as custom sampling procedures for disk lights in two different systems: the Mitsuba renderer~\cite{Jakob:2010:Mitsuba} and the Arnold production renderer~\cite{Fajardo:2010:Arnold}.

In order to sample from each mapping, we need to find the roots of \Eqs{eq:urena_root_finding}{eq:ibon_root_finding} respectively. Since the elliptic integrals they contain prevent analytical inversion, we resort to numerical root finding using an iterative Newton-Raphson method. However, this method can become very expensive, since for each iteration we need to numerically evaluate two and one incomplete elliptic integrals (for the parallel and radial mappings, respectively).

\subsection{Tabulation}

In order to reduce the significant cost of Newton-Raphson over area sampling (up to 10$\times$ in simple scenes; see \Fref{fig:graphs}) and avoid the expensive numerical inversion, we approximate our radial mapping by tabulating \Fref{eq:elliptic_integral}. We choose to tabulate this mapping as it introduces less distortion in the output sample distribution than the parallel one, as shown in \Fref{fig:disks}.


We can write the fractional solid angle $\solidAngleRadial' = \solidAngleRadial(\angleRadial)/\solidAngleRadial(\pi/2)$ as a function of $\alpha \in [0,\pi/2]$, $\beta \in [0,\alpha]$ and
$\angleRadial \in [0,\pi/2]$. We can tabulate this function by discretizing each of the three parameters,
producing a 3D array of values. A quick binary
search based on $\angleRadial$ (combined with interpolation) then allows us to get
approximate values of $\solidAngleRadial'$ with good accuracy. However, storing such a table would require a large
amount of memory. To address this, we reparametrize $\solidAngleRadial'$ in terms of
$\alpha$, the ratio $\beta'=\beta/\alpha$ (which is in $[0,1]$), and $\angleRadial$. This
version of $\solidAngleRadial'$ has very low variation w.r.t. to $\alpha$, so we can remove this parameter
altogether, reducing the tabulation to a 2D array of $\solidAngleRadial'$ values for a set of $\beta'$ and
$\angleRadial$ values. Each entry in this array corresponds to a spherical triangle defined by the 
fraction of $\angleRadial$ covered by the given entry and the value of $\theta(\angleRadial)$ at the 
start of the entry's interval, which can be easily sampled~\cite{Arvo:1995:Triangles}.
This approximation causes some generated samples to lie outside the spherical ellipse, which we 
reject. \new{Note that this rejection ensures unbiasedness; however, for practical reasons our production renderer implementation simply assigns zero weight to such invalid samples, resulting in a slight understimation of the illumination.}
We have found the rejection ratio to be negligible, the storage requirement low,
and the accuracy satisfactory for realistic rendering.
In our implementation we use a 2D table with resolution
$1024^2$, which we found to be accurate enough to provide an insignificant difference in 
variance compared to the analytic solution.
Finally, note that in order to compute the samples' pdf $p(\ww) = 1 / |\lightSolidAngle|$, we still need to compute $\lightSolidAngle$ numerically. This computation is amortized among all samples for a given shading point.

\subsection{Efficiency}

Similarly to existing solid angle sampling techniques (e.g.\ for spherical triangles \cite{Arvo:1995:Triangles} and rectangles \cite{Urena:2013:Quads}), the cost of drawing a sample with our technique is higher than that of uniform area sampling (though some of it is amortized over multiple samples). This overhead pays off when the luminaire is close to the shading point (i.e.\ the subtended solid angle is relatively large).

As an optimization, our Arnold implementation (\Fref{fig:arnold}) employs a simple heuristic to switch to uniform area sampling when the luminaire is far away. In order to provide a fair comparison against existing techniques (in Figures \ref{fig:comparative_surface} and \ref{fig:comparative_media}), our Mitsuba implementation does not take advantage of this optimization.

Even with the above tabulation scheme, we still need to compute the solid angle of the spherical ellipse $\lightSolidAngle$ for the sampling pdf, using either \Fref{eq:urena_int_real} or \eqref{eq:elliptic_integral}. The elliptical integrals involved can be computed using Carlson's fast numerical algorithms \cite{Carlson:1995:EllipticIntegrals}.

\section{Results}
\label{sec:implementation}


\Figs{fig:comparative_surface}{fig:comparative_media} show a comparison between traditional area sampling~\cite{Shirley:1997:Disks}, Gamito's rejection-based solid angle sampling~\cite{Gamito:2016:Disks} and our techniques (Mitsuba implementation), without and with the presence of participating media. In both figures only direct illumination (single scattering) is computed, using 16 samples/pixel. \new{Inside a medium, we first sample a point along the ray via equiangular sampling~\cite{Kulla:2012:Importance} w.r.t.\ the disk light center, and then use the corresponding disk sampling technique to generate a point on the light. In the case of uniform area sampling, a better strategy is to first sample the light surface and then perform equiangular sampling w.r.t.\ that chosen point. We therefore include this strategy in \Fref{fig:comparative_media} (called ``Area sampling (first)''), which is incompatible with the solid angle mappings.} The results show that our sampling methods outperform Gamito's method on surfaces, and perform at least on par in participating media, where variance due to medium sampling dominates when using solid angle sampling. In all cases, area sampling yields much higher variance. In the supplemental document we provide global illumination comparisons between our tabulated radial sampling and Mitsuba's built-in disk area sampling.

\begin{figure}[t!]
	\centering
	\def\svgwidth{\columnwidth}
	\fontsize{0.15cm}{1em}
	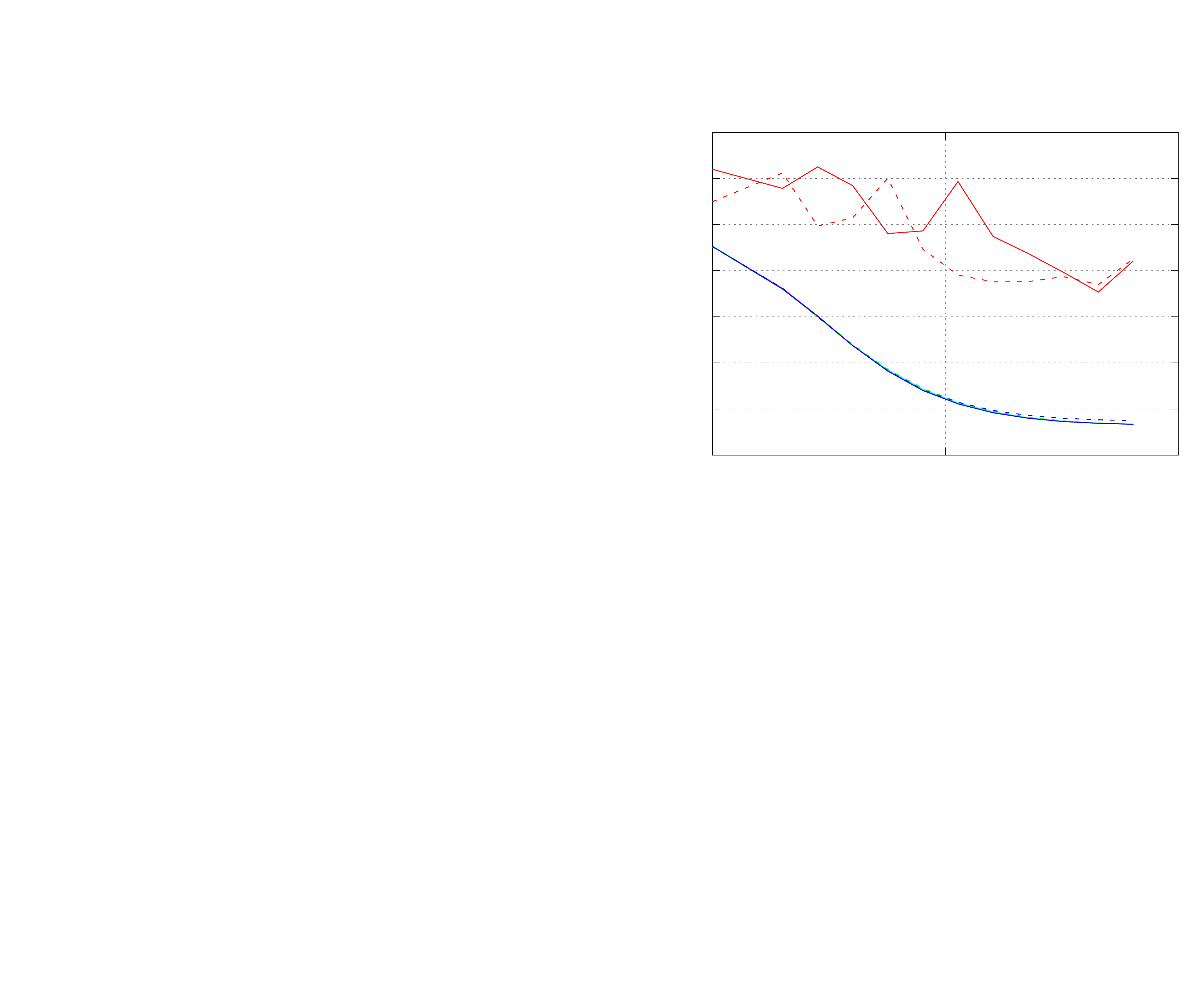
	\vspace{-1.0em}
	\caption{Error ({\itshape top}) and cost ({\itshape bottom}) for the results shown in \Fref{fig:comparative_surface} ({\itshape left}) and \Fref{fig:comparative_media} ({\itshape right}), w.r.t.\ sample count.}
	\label{fig:graphs}
\end{figure}

\Fref{fig:graphs} shows a comparison between the convergence and the cost of the different techniques from \Figs{fig:comparative_surface}{fig:comparative_media}. For the same number of samples, our mappings produce lower error than Gamito on surfaces and perform virtually identically in media.
In terms of cost, our tabulated version is almost as fast as area sampling, and the fully numerical implementation can be up to 10$\times$ slower. Note that the performance of the iterative numerical inversion depends on the geometrical configuration: the starting point for the inversion affects the number of iterations required for convergence.
The parallel and radial mappings take respectively 1-3 and 1-4 Newton-Raphson iterations in our tests. Also note that in scenes with higher geometric and shading complexity, the relative cost of the different methods has less impact on the overall rendering performance.

\begin{figure*}[h!]
	\centering
	\includegraphics[trim={0 1.4cm 0 5cm},clip,width=1.044\columnwidth]{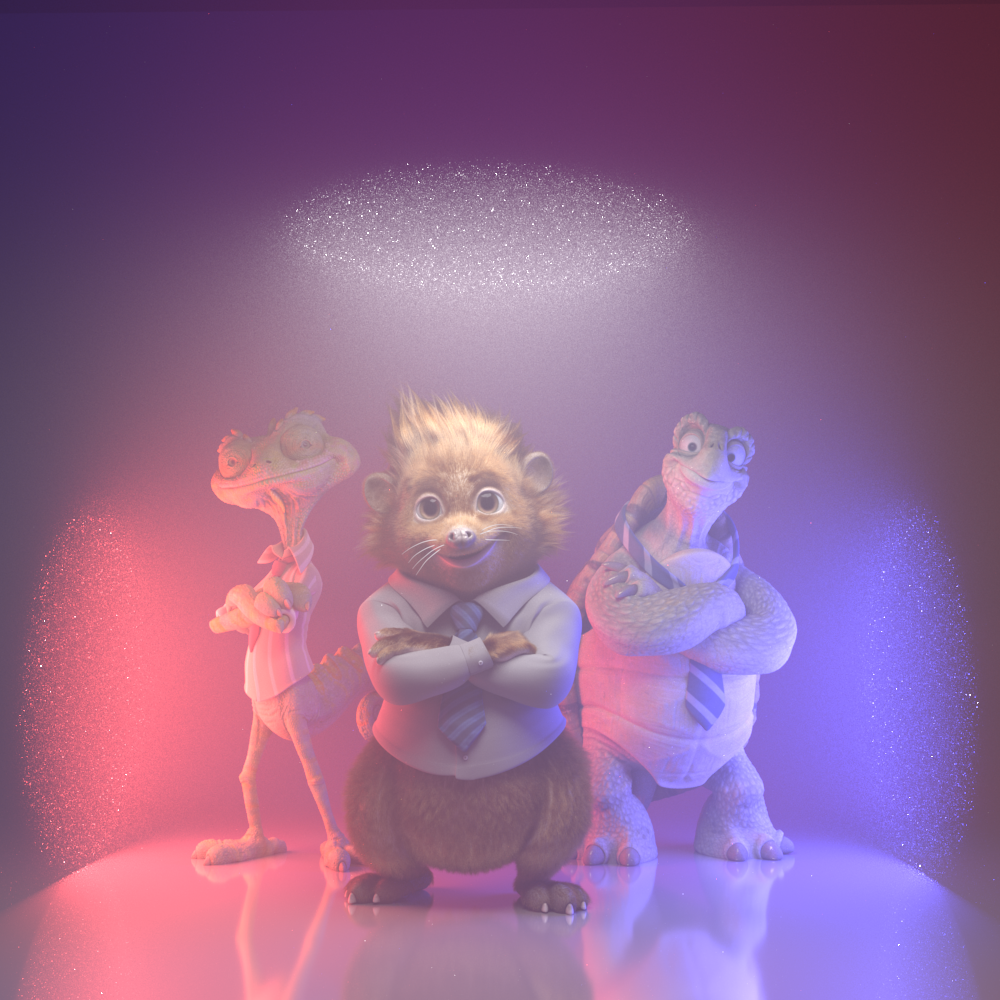}
	\includegraphics[trim={0 1.4cm 0 5cm},clip,width=1.044\columnwidth]{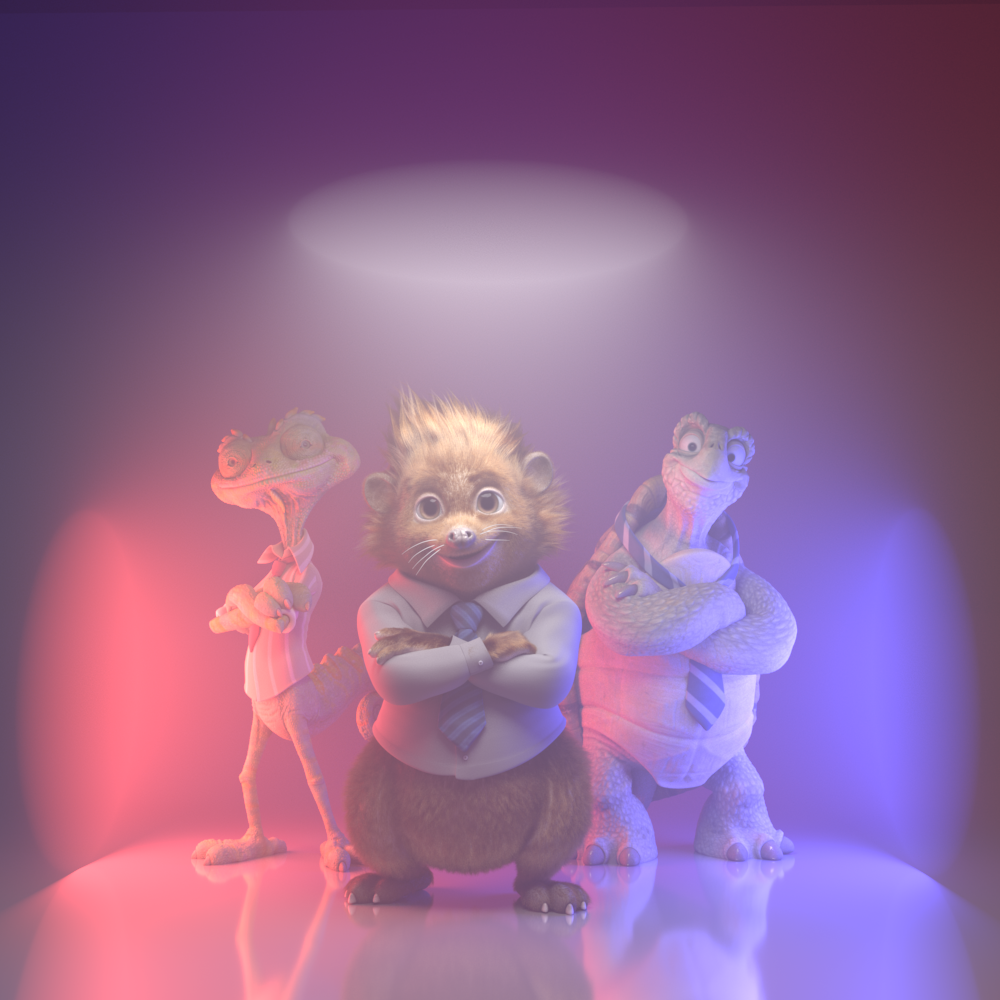}
\caption{Colored disk lights (invisible) rendered in Arnold using area sampling ({\itshape left}) and our tabulated radial map ({\itshape right}) with 256 samples/pixel. Due to the use of complex surface and hair shaders, the higher cost of our technique has a negligible impact on performance.
\label{fig:arnold}
}
\end{figure*}

Finally, \Fref{fig:arnold} shows a scene rendered in Arnold, comparing our tabulated radial map implementation to the renderer's built-in uniform area sampling. The scene features many production features, including high-resolution texture maps, fur, displacement, subsurface scattering, indirect surface and volume-to-surface light transport. In such cases the higher cost of our technique has a negligible impact on the total rendering performance. With 256 samples/pixel our tabulated radial map yields a noise-free image, while area sampling suffers from a substantial amount of noise.

\section{Conclusions}
We have presented two new area-preserving mappings that enable the uniform solid angle sampling of oriented disk light sources.
Following the key observation that this solid angle is a spherical ellipse, we make use of the Hat-Box theorem to transform canonical unit-square sample points onto the spherical ellipse in a way that preserves their stratification. To avoid costly numerical inversion, we develop a practical mapping tabulation that introduces little overhead over traditional uniform area sampling~\cite{Shirley:1997:Disks} while significantly reducing the variance of the illumination estimate. Our mappings are also competitive to existing disk solid angle sampling techniques~\cite{Gamito:2016:Disks}, without imposing restrictions on the sample generator. 

As a by-product of our work, we have proposed two new expressions for the subtended solid angle of a disk, which in addition to graphics is important in other fields such as particle transport. In this context, most previous analytic formulations~\cite{Paxton:1959:SA,Timus:2007:FARSA,Conway:2010:ASA} have included at least two incomplete elliptic integrals that need to be computed numerically. In contrast, our radial formulation, based on Booth's spherical topology analysis~\cite{Booth:1844:SphericalEllipse}, involves only one elliptic integral, making it more simple and practical than previous work. 

While this work only considers circular disks, our approach could be extrapolated to other shapes whose subtended solid angle is also an ellipse, such as elliptical disks and ellipsoids~\cite{Heitz2017}. Including these geometries would only require finding the spherical ellipses subtended by them. Moreover, our mappings could reduce variance of other shapes such as cylinders, following Gamito~\cite{Gamito:2016:Disks}. 

The main limitation of our method is the lack of analytical inversion of the proposed mappings, which requires using either costly numerical inversion or tabulation. Unfortunately, it seems impossible to find a spherical ellipse mapping that does not involve incomplete elliptic integrals, whose inversion is unknown.
The presented mappings also only consider the solid angle, but not the other contribution terms in \Fref{eq:rendering}, e.g.\ the BRDF or the foreshortening term. Developing methods for including at least some of these other terms is an interesting direction for future work. \new{Furthermore, while our mappings are nearly optimal for uniformly emitting disk luminaires, it would be interesting to take into account spatially-varying emission profiles, in the spirit of the work of Bitterli et al.~\cite{Bitterli:2015:Portal}.}

\renewcommand{\fithlinewidth}{0.196\textwidth}
\newcommand{\samplesSurface}{16}
\newcommand{\sceneSurface}{spheres_bottom}
\begin{figure*}[!ht]
	{
	\centering
	\begin{subfigure}[t]{\fithlinewidth}
		\includegraphics[width=\linewidth]{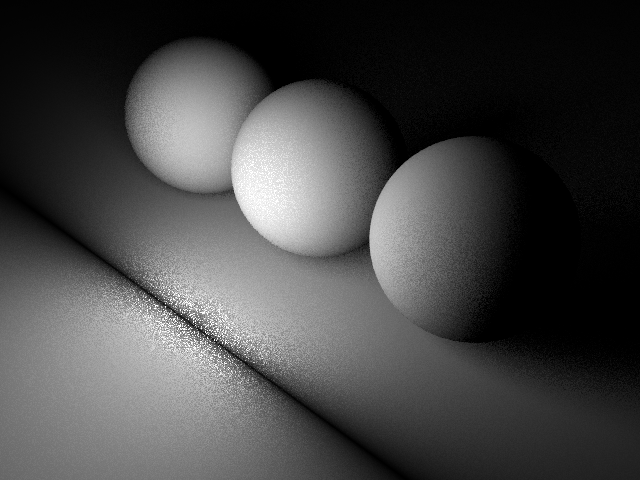}
		\includegraphics[width=\linewidth]{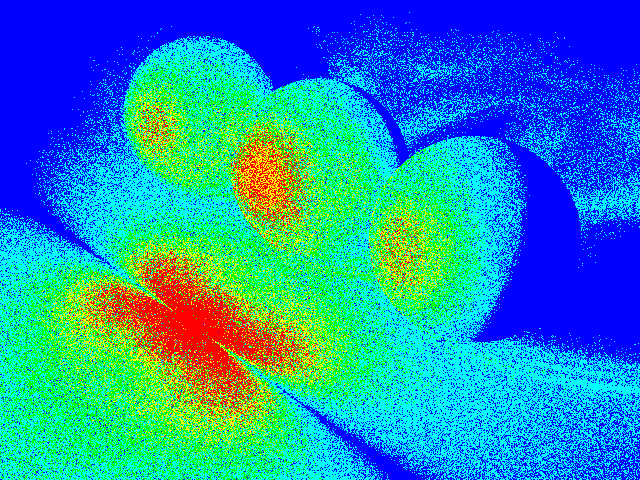}
		\captionsetup{justification=centering}
		\subcaption{Area sampling~\cite{Shirley:1997:Disks}}
	\end{subfigure}
	\begin{subfigure}[t]{\fithlinewidth}
		\includegraphics[width=\linewidth]{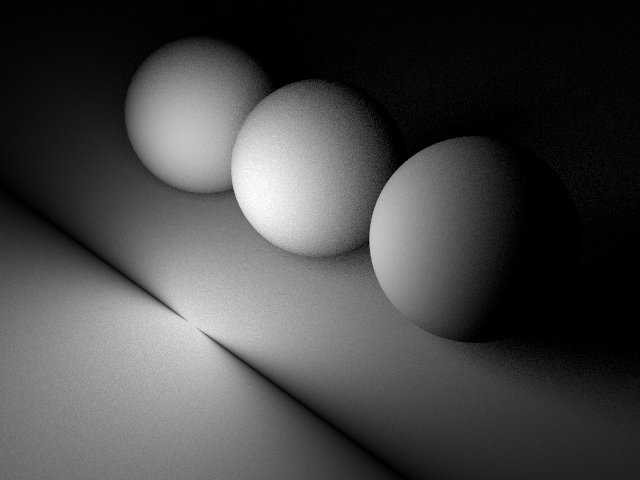}
		\includegraphics[width=\linewidth]{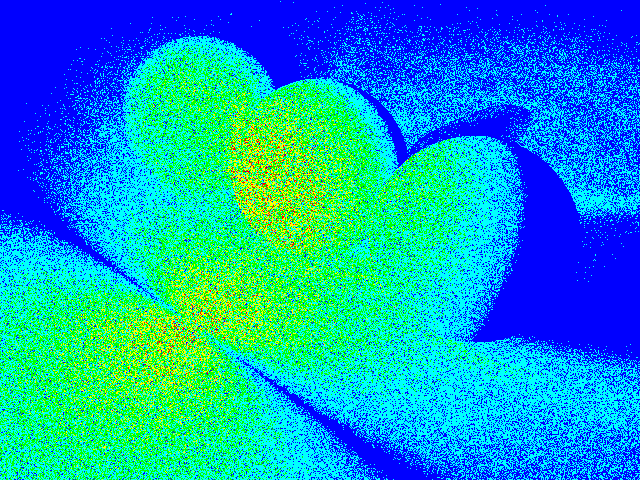}
		\captionsetup{justification=centering}
		\subcaption{Gamito~\cite{Gamito:2016:Disks}}
	\end{subfigure}
	\begin{subfigure}[t]{\fithlinewidth}
		\includegraphics[width=\linewidth]{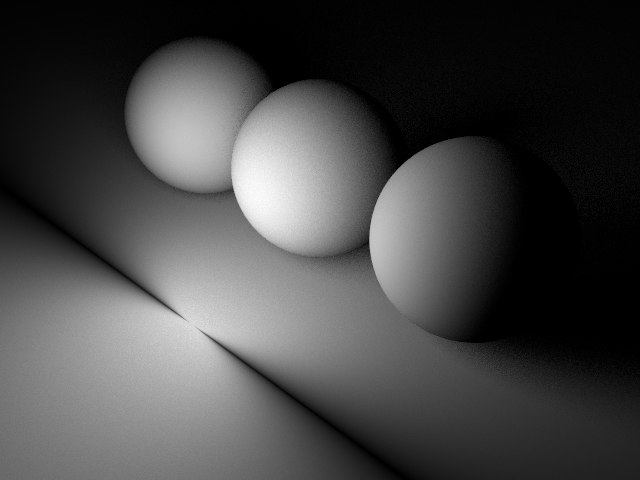}
		\includegraphics[width=\linewidth]{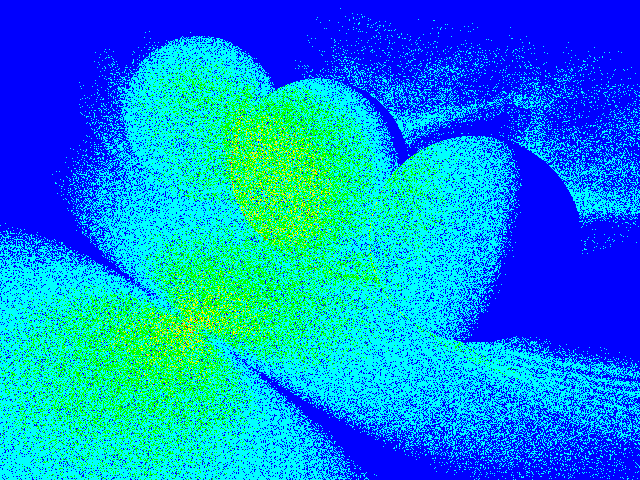}
		\captionsetup{justification=centering}
		\subcaption{Our parallel mapping}
	\end{subfigure}
	\begin{subfigure}[t]{\fithlinewidth}
		\includegraphics[width=\linewidth]{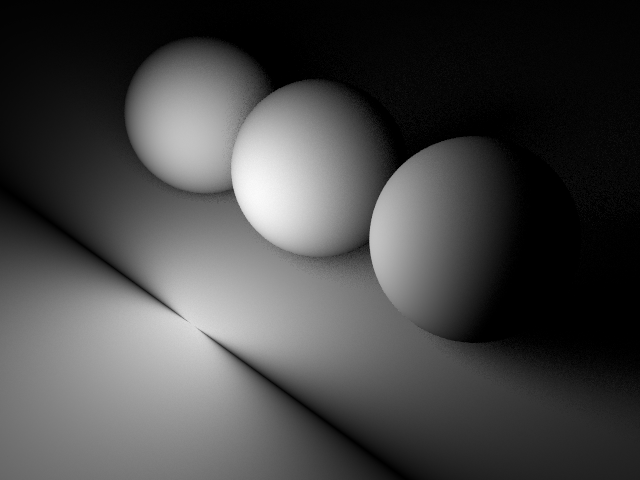}
		\includegraphics[width=\linewidth]{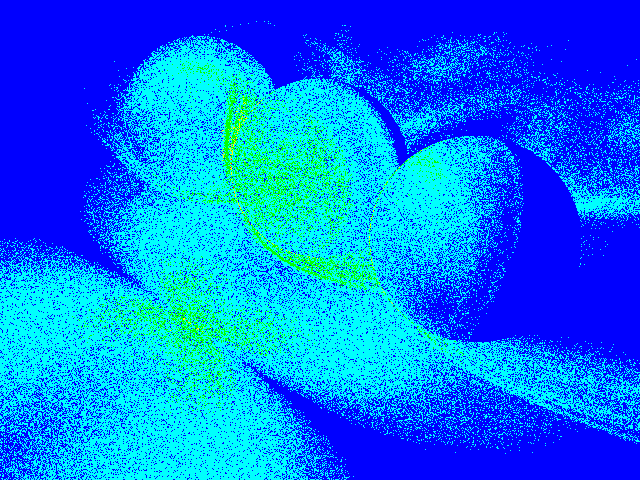}
		\captionsetup{justification=centering}
		\subcaption{Our radial mapping}
	\end{subfigure}
	\begin{subfigure}[t]{\fithlinewidth}
		\includegraphics[width=\linewidth]{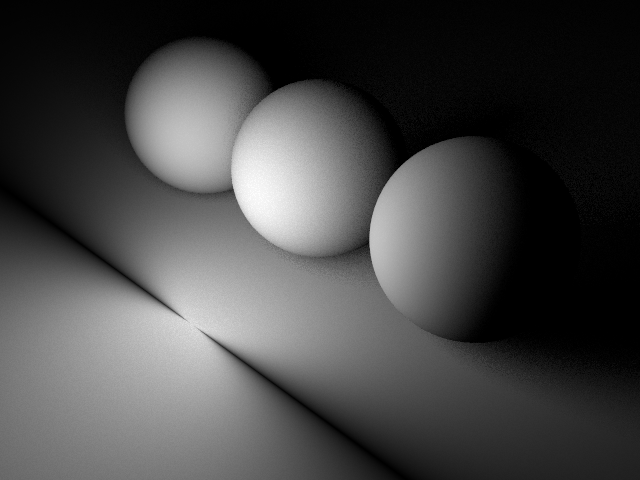}
		\includegraphics[width=\linewidth]{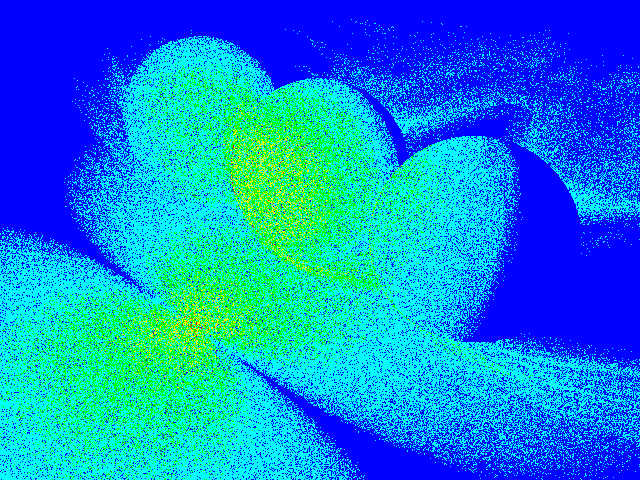}
		\captionsetup{justification=centering}
		\subcaption{Our ld-radial mapping}
	\end{subfigure}
	}
	\begin{picture}(0,0)%
		\put(28.4,84.5){\color[rgb]{1,1,1}\makebox(0,0)[rb]{\tiny{$0.0043451$}}}%
		\put(139.5,83.8){\color[rgb]{1,1,1}\makebox(0,0)[rb]{\tiny{$3.7234 \!\!\times\!\! 10^{-5}$}}}%
		\put(240.5,83.8){\color[rgb]{1,1,1}\makebox(0,0)[rb]{\tiny{$1.2195 \!\!\times\!\! 10^{-5}$}}}%
		\put(342.2,83.8){\color[rgb]{1,1,1}\makebox(0,0)[rb]{\tiny{$3.1013 \!\!\times\!\! 10^{-6}$}}}%
		\put(443.3,83.8){\color[rgb]{1,1,1}\makebox(0,0)[rb]{\tiny{$1.2655 \!\!\times\!\! 10^{-5}$}}}%
	\end{picture}%
        \vspace{-0.5em}
	\caption{
	{\itshape Top}: Scene illuminated by a double-sided disk light, rendered with \samplesSurface{} samples/pixel. The light is perpendicular to the ground and is invisible to camera rays.
	{\itshape Bottom}: False-color differences and MSE w.r.t.\ to a reference image computed with 32K samples/pixel.
	}
	\label{fig:comparative_surface}
\end{figure*}
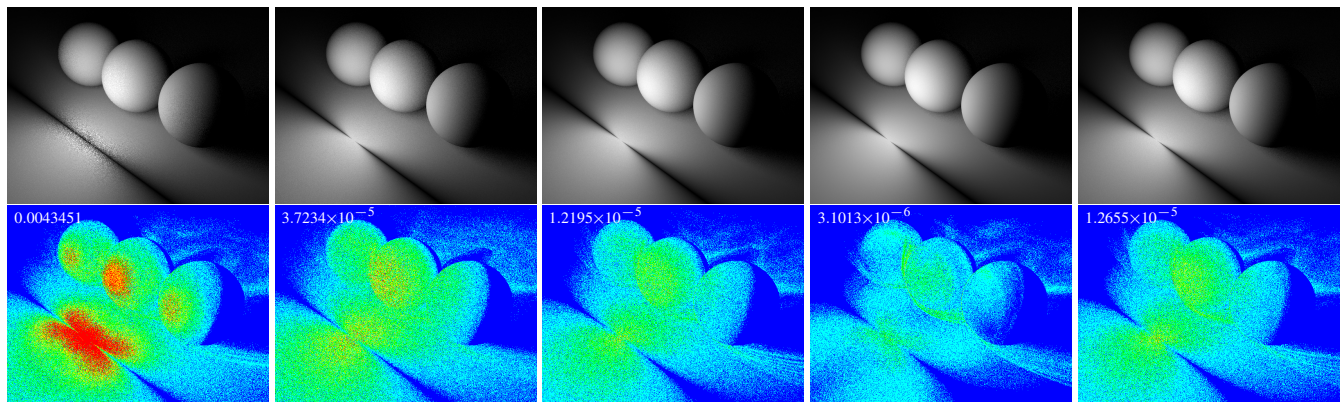

\renewcommand{\fithlinewidth}{0.163\textwidth}
\newcommand{\samplesMedia}{16}
\newcommand{\sceneMedia}{bunny_media4}
\begin{figure*}[!ht]
	{
	\centering
	\begin{subfigure}[t]{\fithlinewidth}
		\includegraphics[width=\linewidth]{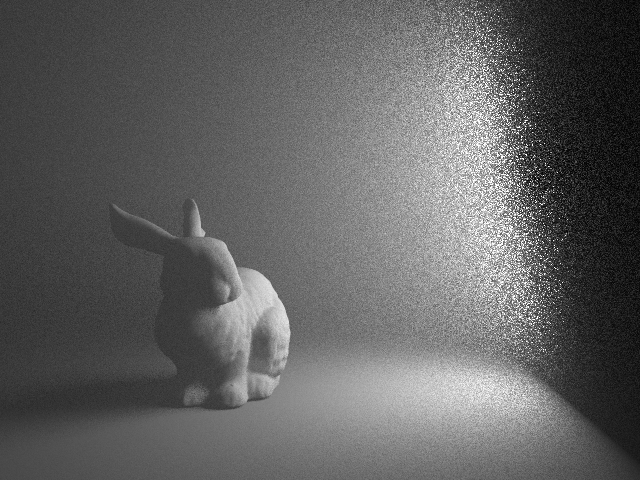}
		\includegraphics[width=\linewidth]{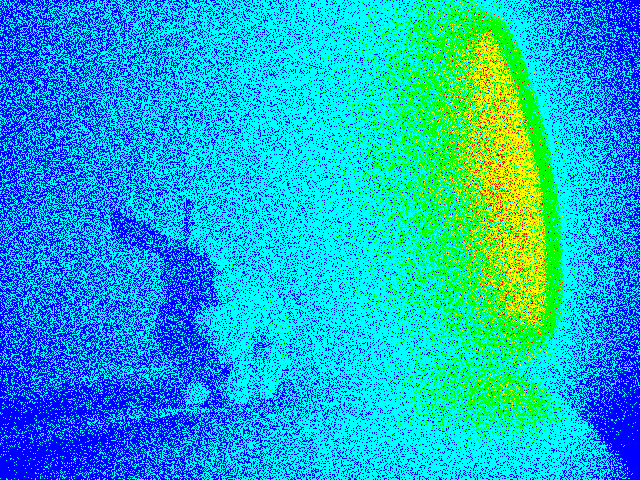}
		\captionsetup{justification=centering}
		\subcaption{Area sampling~\cite{Shirley:1997:Disks}}
	\end{subfigure}
	\begin{subfigure}[t]{\fithlinewidth}
		\includegraphics[width=\linewidth]{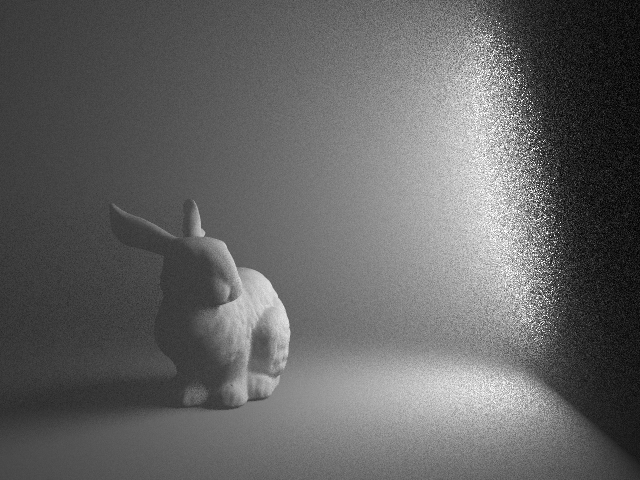}
		\includegraphics[width=\linewidth]{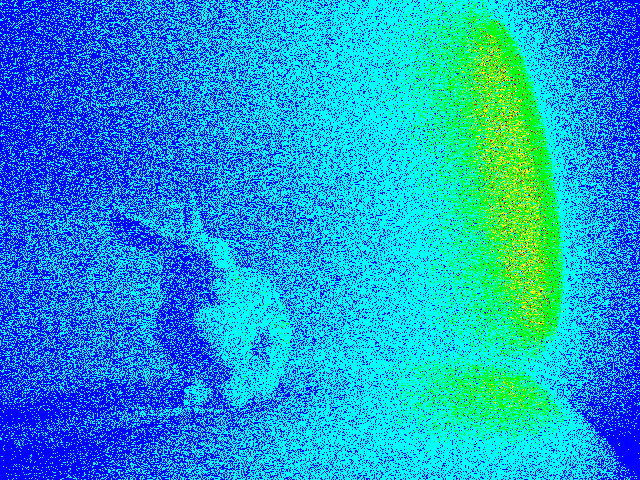}
		\captionsetup{justification=centering}
		\subcaption{Area sampling (first)}
	\end{subfigure}
	\begin{subfigure}[t]{\fithlinewidth}
		\includegraphics[width=\linewidth]{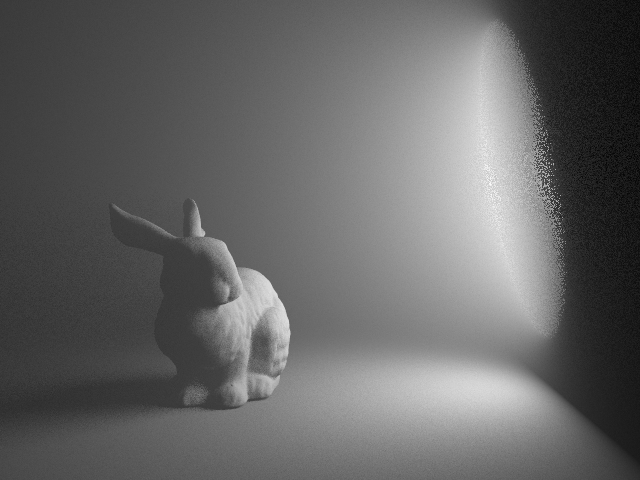}
		\includegraphics[width=\linewidth]{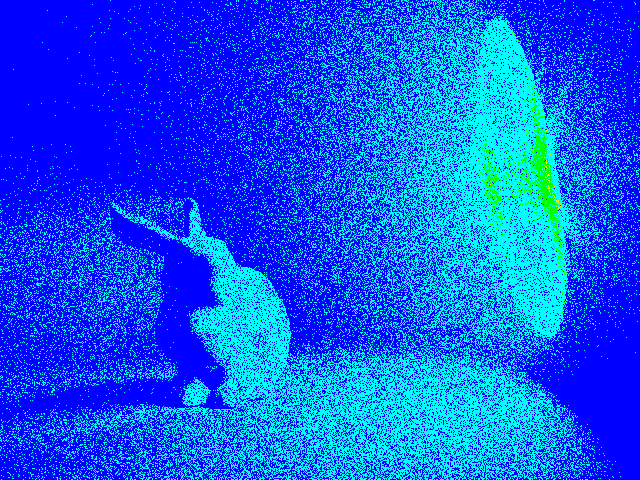}
		\captionsetup{justification=centering}
		\subcaption{Gamito~\cite{Gamito:2016:Disks}}
	\end{subfigure}
	\begin{subfigure}[t]{\fithlinewidth}
		\includegraphics[width=\linewidth]{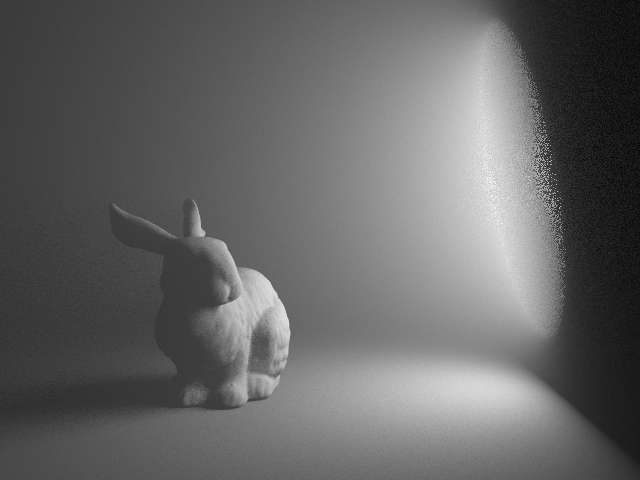}
		\includegraphics[width=\linewidth]{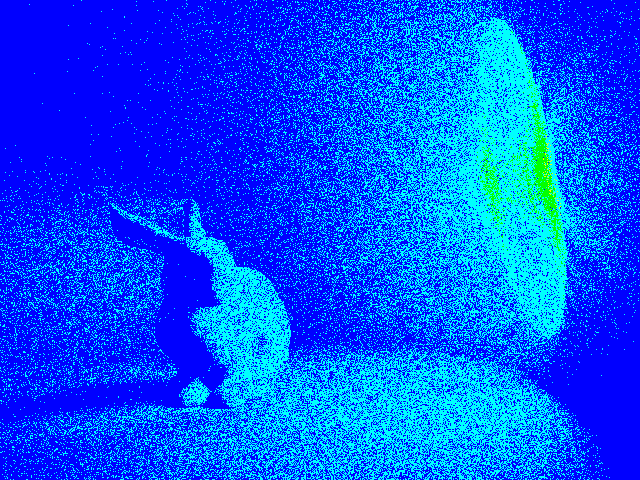}
		\captionsetup{justification=centering}
		\subcaption{Our parallel mapping}
	\end{subfigure}
	\begin{subfigure}[t]{\fithlinewidth}
		\includegraphics[width=\linewidth]{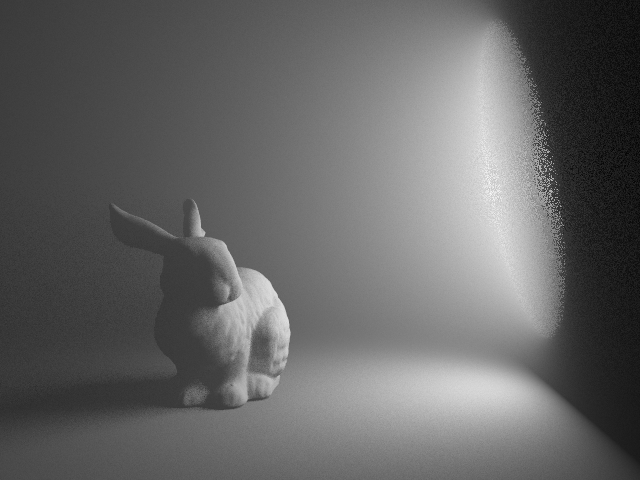}
		\includegraphics[width=\linewidth]{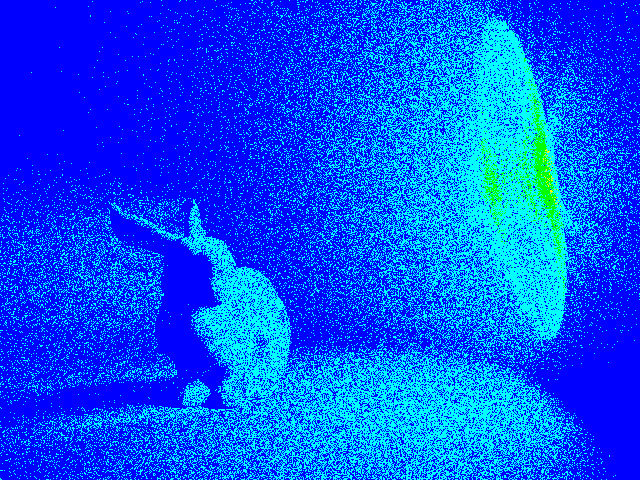}
		\captionsetup{justification=centering}
		\subcaption{Our radial mapping}
	\end{subfigure}
	\begin{subfigure}[t]{\fithlinewidth}
		\includegraphics[width=\linewidth]{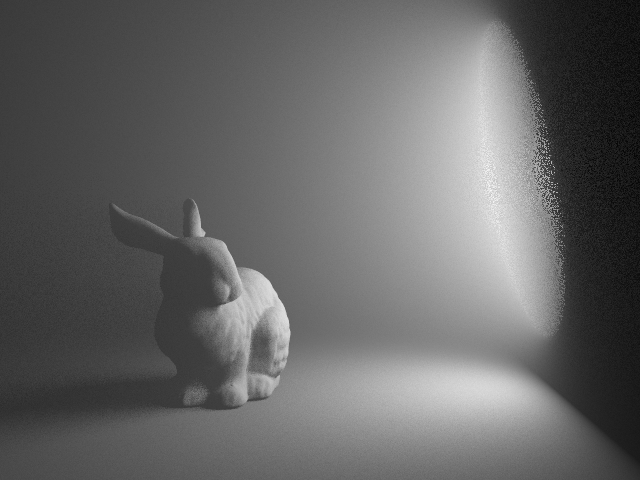}
		\includegraphics[width=\linewidth]{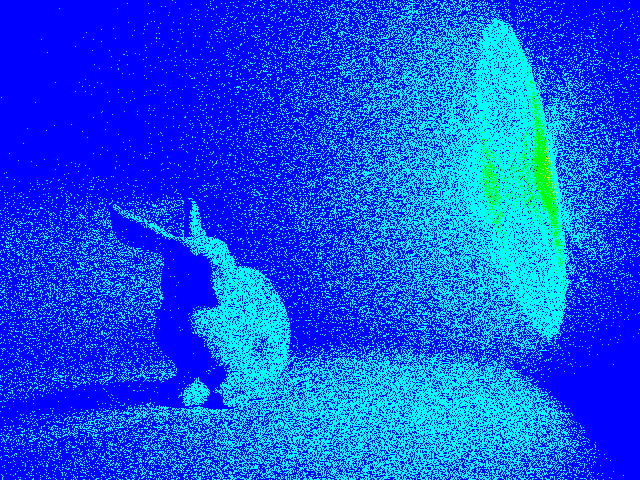}
		\captionsetup{justification=centering}
		\subcaption{Our ld-radial mapping}
	\end{subfigure}
	}
	\begin{picture}(0,0)%
		\put(22.3,72.2){\color[rgb]{1,1,1}\makebox(0,0)[rb]{\tiny{$0.70019$}}}%
		\put(107,72.2){\color[rgb]{1,1,1}\makebox(0,0)[rb]{\tiny{$0.14059$}}}%
		\put(207,71.5){\color[rgb]{1,1,1}\makebox(0,0)[rb]{\tiny{$2.4044 \!\!\times\!\! 10^{-4}$}}}%
		\put(291,71.5){\color[rgb]{1,1,1}\makebox(0,0)[rb]{\tiny{$2.3957 \!\!\times\!\! 10^{-4}$}}}%
		\put(376,71.5){\color[rgb]{1,1,1}\makebox(0,0)[rb]{\tiny{$2.3610 \!\!\times\!\! 10^{-4}$}}}%
		\put(460,71.5){\color[rgb]{1,1,1}\makebox(0,0)[rb]{\tiny{$2.3739 \!\!\times\!\! 10^{-4}$}}}%
	\end{picture}%
        \vspace{-0.5em}
	\caption{
	{\itshape Top}: A scene with a participating medium illuminated by a single-sided disk light (invisible), rendered with \samplesMedia{} samples/pixel. Please refer to Section~\ref{sec:implementation} for details.
	{\itshape Bottom}: False-color differences and MSE w.r.t.\ to a reference image computed with 64K samples/pixel.
	}
	\label{fig:comparative_media}
\end{figure*}
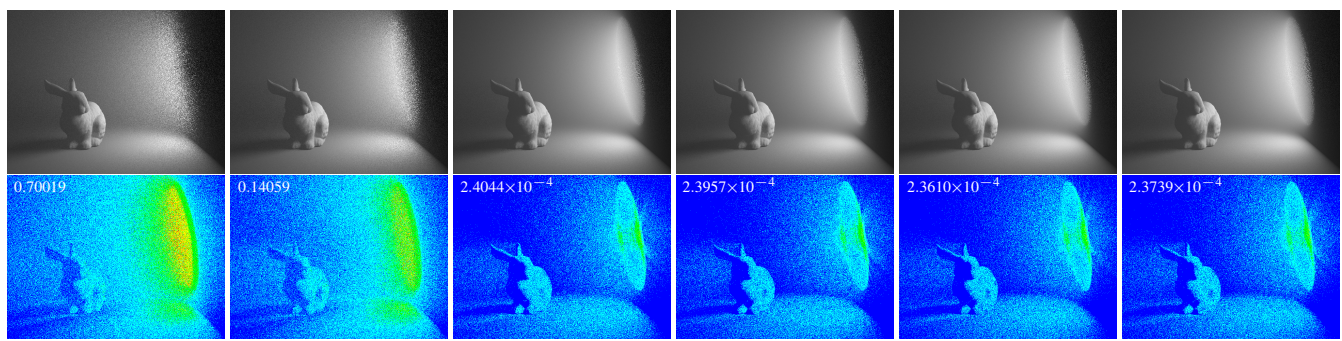

\section*{Acknowledgements}
We would like to thank the anonymous reviewers for their suggestions. This project was funded by the European Research Council (ERC Consolidator Grant 682080), DARPA (HR0011-16-C-0025), and the Spanish Ministerio de Econom\'{i}a y Competitividad (TIN2014-61696-EXP, TIN2013-47276-C6-3-R). Jorge L\'{o}pez-Moreno was additionally founded by a Juan de la Cierva fellowship. No ellipses were harmed in the making of this paper.


\bibliographystyle{eg-alpha-doi}

\bibliography{src_bib}


\newcommand{\atangsqi}{q}
\newcommand{\btangsqi}{p}

\newpage

\appendix
\section{Derivation of \Fref{eq:delta_sin}}
\label{sec:h1deriv}

\begin{figure}[b!]
   	\centering
    \def\svgwidth{0.5\linewidth}
    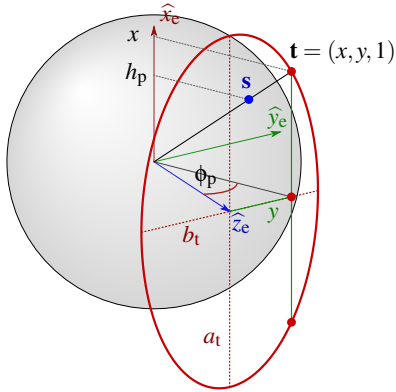
    \vspace{-0.5em}
   	\caption{A view of the tangent ellipse (red).
   	For a given angle $\anglePolar$, we first obtain the coordinate $y = \tan\anglePolar$ along axis $\axis{y}[e]$. Using the ellipse equation, we then find the corresponding coordinate $x$ along $\axis{x}[e]$. This gives point $\point{t}$ on the tangent ellipse, whose spherical projection $\point{s}$ has $\axis{x}[e]$-coordinate $\hPolar$ -- the quantity we are interested in.
   	}
   	\label{fig:tangent-ellipse}
   	\vspace{-0.5em}
\end{figure}

Here we derive the expression for $\hPolar(\anglePolar)$ in \Fref{eq:delta_sin}, whose integral we then express as a combination of incomplete elliptic integral functions in \Fref{eq:urena_int_real}. We use the tangent ellipse, shown in \Fref{fig:tangent-ellipse} and introduced in \Fref{sec:desarrollo} and \Fref{fig:projection}, right. The ellipse semi-axes $\atang \geq \btang$ are aligned with $\axis{x}[e]$ and $\axis{y}[e]$, respectively.

For any angle $\anglePolar \in [-\beta,\beta]$, we first obtain a coordinate $y = \tan\anglePolar$ along the $\axis{y}[e]$ axis. (We only consider $\anglePolar > 0$, thus $y>0$, and convert negative $\anglePolar$ to positive using symmetry, as described in \Fref{sec:urena_mapping}.) Using the ellipse equation $(x / \atang)^2 + (y / \btang)^2 = 1$, we can get the corresponding $x \geq 0$ coordinate along $\axis{x}[e]$ as a function of $y$:
\begin{equation}\label{eq:app_y}
      x ~=~\atang\,\sqrt{1-(y/\btang)^2}.
\end{equation}
We then consider the point $\point{t}=(x,y,1)$ on the tangent ellipse and its spherical projection $\point{s}=\point{t}/\|\point{t}\|$.
The $\axis{x}[e]$-coordinate of $\point{s}$, and also of its cylindrical projection (see \Fref{fig:urena_integration}, left), is
\begin{equation}
    \hPolar = \frac{x}{\sqrt{x^2+y^2+1}}.
    \label{eq:h1}
\end{equation}
Substituting \Fref{eq:app_y} into \eqref{eq:h1}:
\begin{equation}\label{eq:app_h1}
  \hPolar = \atang\,
     \frac{\sqrt{1-\btangsqi y^2}}
          {\sqrt{y^2+1+\atang^2\left(1-\btangsqi y^2\right)}}
          = \ctang\,\sqrt{\frac{1-p\,y^2}
                            {1-m\,p\, y^2}}
\end{equation}
where $p$, $m$ and $\ctang$ are as in \Fref{eq:pmc}.

Using $y = \tan\anglePolar$ and $0\leq \anglePolar\leq \beta\leq \pi/2$, in \Fref{eq:app_h1} we can substitute $y^2$ by $(\sin^2\anglePolar)/(1-\sin^2\anglePolar)$. With this we can finally write $\hPolar$ explicitly as a function of $\anglePolar$:
\begin{equation}
   \hPolar(\anglePolar)
          ~ = ~ \ctang\,\sqrt{\frac{1-(p+1)\sin^2\anglePolar}
                               {1-(mp+1)\sin^2\anglePolar}}
\end{equation}
which is exactly \Fref{eq:delta_sin}.

\section{Derivation of \Fref{eq:ellipse radius}}
\label{sec:deriv_ellipse_radius}

\newcommand{\pointr}{\point{r}}

Here we derive the expression for $r(\angleRadial)$  in \Fref{eq:ellipse radius}, which is used in the radial mapping (\Fref{sec:ibon_mapping}). We consider the planar ellipse resulting from the parallel projection of the spherical ellipse onto the $\axis{x}[e]\axis{y}[e]$ plane. This ellipse's semi-major and semi-minor axes are $a$ and $b$, respectively (see \Fref{fig:ellipse_quadrant}). 

Consider a point $\pointr$ whose polar coordinates $(\angleRadial,r(\angleRadial))$ and Cartesian coordinates $(x,y)$ are related as
\begin{equation}
    x \,=\, r(\angleRadial)\cos\angleRadial, ~~~~~~~~
    y \,=\,r(\angleRadial)\sin\angleRadial.
    \label{eq:coord_relation}
\end{equation}
We want to define $r(\angleRadial)$ in such a way that $\pointr$ is on the planar ellipse curve. Thus, $x$ and $y$ must obey the ellipse equation, i.e.\
\begin{equation}\label{eq:pointr elleq}
   \left(\frac{x}{a}\right)^2 + \left(\frac{y}{b}\right)^2 \, = \, 1.
\end{equation}
We can substitute $x$ and $y$ from \Fref{eq:coord_relation} into the ellipse equation, resulting in
\begin{equation}
   r(\angleRadial)^2\,
   \left(
      \frac{b^2\cos^2\angleRadial}{b^2 a^2} +
      \frac{a^2\sin^2\angleRadial}{a^2 b^2}
   \right)
   \,=\,1.
\end{equation}
We can thus write
\begin{equation}
   r(\angleRadial) \,=\,
   \frac{ab}{\sqrt{b^2\cos^2\angleRadial + a^2\sin^2\angleRadial}},
\end{equation}
which is  \Fref{eq:ellipse radius}.

\end{document}